\documentclass[fleqn,usenatbib]{mnras}

\usepackage[T1]{fontenc}
\usepackage{ae,aecompl}

\usepackage{graphicx}	
\usepackage{amsmath}	
\usepackage{amssymb}	

\makeatletter
\newlength{\abovecaptionskip}%
\makeatother
\usepackage{threeparttable}%

\usepackage{newtxtext,newtxmath}

\title[VLA monitoring of JVAS~B1422+231]{A VLA monitoring study of JVAS~B1422+231: investigation of time delays and detection of extrinsic variability}

\author[A. D. Biggs]{
  A.~D.~Biggs$^{1,2}$\thanks{E--mail: andrew.biggs@stfc.ac.uk}
  \\
  $^{1}$European Southern Observatory, Karl Schwarzschild Stra{\ss}e 2, D-85748 Garching, Germany\\
  $^{2}$UK Astronomy Technology Centre, Royal Observatory, Blackford Hill, Edinburgh EH9 3HJ
}

\date{Accepted XXX. Received YYY; in original form ZZZ}

\pubyear{2023}

\begin{document}
\label{firstpage}
\pagerange{\pageref{firstpage}--\pageref{lastpage}}
\maketitle

\begin{abstract}
  We present an analysis of two seasons of archival, multi-frequency VLA monitoring of the quad lens system JVAS~B1422+231, the 15-GHz data of which have previously been published. The 8.4- and 15-GHz variability curves show significant variability, especially in polarization, but lack features on short time-scales that would be necessary for an accurate measurement of the very short predicted time delays ($\la$1~d) between the three bright images. Time delays can only realistically be measured to the very faint image~D and for the first time we detect its long-term variability and determine its polarization properties. However, image-dependent (extrinsic) variability (including variations on time-scales of hours) is present in multiple images and the magnitude of this is largest in image~D at 15~GHz ($\pm$10~per~cent). As the variations appear to increase in amplitude with frequency, we suggest that the most likely cause is microlensing by compact objects in the lensing galaxy. Combining the monitoring data allows us to detect a faint arc of emission lying between images B and C and the jets responsible for this are imaged using archival VLBA data. Finally, we have also measured the rotation measure of the three bright images and detected the polarization properties of image~D.
\end{abstract}




\section{Introduction}
\label{sec:intro}

\defcitealias{patnaik01b}{PN01}

JVAS~B1422+231 \citep{patnaik92b} is a four-image gravitational lens system discovered as part of the Jodrell Bank/VLA Astrometric Survey \citep[JVAS -- ][]{patnaik92}. A so-called `cusp' lens \citep*[e.g.][]{keeton03a}, three of the images (A, B and C) are bright ($\ga 100$~mJy at 8.4~GHz) and approximately collinear whilst the fourth is more than thirty times fainter than the brightest image and lies close to and on the other side of the lensing galaxy. The maximum angular separation is $\sim$1.3~arcsec. The background source is a $z = 3.62$ quasar and the lensing galaxy is an elliptical \citep{yee94,impey96} that forms part of a group at $z = 0.338$ \citep{kundic97a,tonry98}.

One of the most interesting aspects of this system is that smooth lens models cannot reproduce the observed flux ratios. This became apparent during the first attempts to model the system \citep*{hogg94,kormann94} and led \citet{mao98} to propose that the flux anomalies were due to substructure in the lensing galaxy, a possibility subsequently investigated in more detail by \citet{bradac02}. Cold Dark Matter ($\Lambda$CDM) models of large-scale structure formation predict that high-mass galaxies should have large numbers of subhaloes \citep[e.g.][]{moore99} and strong gravitational lensing can be used to investigate the substructure mass function of high-redshift galaxies, either by analysis of the flux ratios \citep[e.g.][]{metcalf01,dalal02} or by directly imaging distortions of lensed emission \citep[e.g.][]{vegetti09,vegetti10}.

\begin{table*}
  \centering
  \begin{threeparttable}
    \caption{Summary of the two VLA monitoring seasons of B1422+231. The AP282 epochs were observed as part of the JVAS observations and contain multiple scans at different hour angles.}
    \label{tab:obs}
    \begin{tabular}{cccccc} \\ \hline
      Season & Project code & Dates & Configurations & Frequency bands (GHz) & Number of epochs \\ \hline
      1 & AP282 & 1994 Feb 20 -- 1994 Feb 25 & A\tnote{\textit{a}} & 8.4, 15, 22\tnote{\textit{b}} & 4 \\
      & AP263 & 1994 Mar 3 -- 1994 Sep 16 & A, B & 8.4, 15 & 49\tnote{\textit{c}} \\
      2 & AH593 & 1996 Oct 10 -- 1997 May 24 & A, B & 8.5 & 44 \\ \hline
    \end{tabular}
    \begin{tablenotes}
    \item [\textit{a}] These observations contained a small number of antennas still in their positions from the preceding D configuration.
    \item [\textit{b}] A single 22-GHz scan was taken during one of these epochs.
    \item [\textit{c}] 51 epochs are in the archive, but only 49 contain all sources. Only 47 calibratable epochs were observed at 15~GHz.
    \end{tablenotes}
  \end{threeparttable}
\end{table*}

Alternatively, it could be that the image flux ratios are being distorted by independent brightness changes in each image. In the optical, this is commonly observed as microlensing by stars in the lensing galaxy, but it was initially assumed that at radio wavelengths the size of the emitting region would render any such variations undetectable \citep[e.g.][]{wambsganss98}. However, Very Large Array (VLA) monitoring later found incontrovertible evidence of image-dependent variability in CLASS~B1600+434 \citep{koopmans00b,biggs21}, whilst two other radio lenses, CLASS~B2045+265 \citep{koopmans03} and JVAS~B1030+074 \citep{rumbaugh15,biggs18b}, also show strong evidence of so-called extrinsic variability. The MERLIN 5-GHz monitoring campaign presented by \citet{koopmans03} also included B1422+231 data, but it was not possible to put strong constraints on whether image-dependent variability was present.

B1422+231 has also been monitored using the VLA and \citet[][hereafter \citetalias{patnaik01b}]{patnaik01b} published time delays between the three bright images using data taken at 15~GHz. If the time delays between the multiple images of a gravitational lens can be determined, a distance-ladder-independent measurement of the Hubble parameter, $H_0$, becomes possible providing that the redshifts of the lensed source and lens galaxy are known and a model for the gravitional deflection of the lens galaxy is available \citep{refsdal64}. The lens model is usually the hardest thing to get right and quad lens are generally preferred for this work as they provide more constraints than two-image systems. However, measuring time delays in B1422+231 should be very difficult given that lens models \citep{hogg94,kormann94,raychaudhury03} predict very short delays between the three bright images, the largest being $\la 1$~d between images B and C, with C leading. However, \citetalias{patnaik01b} measure $\tau_{\mathrm{B-C}} = 8.2 \pm 2.0$~d, a value that is more than $3~\sigma$ higher than expected.

The inconsistency between the predicted and measured delays has encouraged us to revisit the monitoring data, not all of which have been published. \citetalias{patnaik01b} did not use their 8.4-GHz data taken during the smaller B~configuration as the angular resolution was felt to be insufficient to reliably measure the image flux densities, but this is partially a consequence of measuring flux densities in the image plane. Our \textit{u,v} modelfitting approach, as well as avoiding various sources of error related to the imaging process e.g.\ gridding and deconvolution errors, is also able to reliably measure the flux densities of source components even when these are separated by less than the size of the synthesized beam \citep[e.g.][]{martividal14}. A further season of 8.4-GHz monitoring is stored in the VLA archive, as well as four epochs taken in the weeks preceding the \citetalias{patnaik01b} monitoring.

Finally, \citetalias{patnaik01b} did not analyse the polarization properties of the images, despite the lensed source being linearly polarized at the few-per-cent level \citep{patnaik92b}. Polarization data potentially allow the measurement of a delay in both polarized flux and polarization position angle and it may be possible to determine a time delay from the polarization data even if the total flux density varies little \citep{biggs18b}. Polarization variability is also usually greater than that of the total intensity \citep{saikia88}.

In this paper we therefore present a full analysis of all VLA monitoring data for B1422+231, including measurements of the polarization properties of the images. This is the latest in a sequence of papers which has looked at archival data on the lens systems JVAS~B0218+357, JVAS~B1030+074 and CLASS~B1600+434. In Section~\ref{sec:obs} we present details of the observations and calibration and in Section~\ref{sec:rlc} we show the resulting radio light curves. A time-delay analysis using the total flux density data is performed in Section~\ref{sec:delay} and in Section~\ref{sec:discussion} we investigate the extrinsic variability of the light curves, the presence of faint emission between the three bright images and their polarization properties. Conclusions are presented in Section~\ref{sec:conclusions}.

\section{Observations and data reduction}
\label{sec:obs}

The first VLA monitoring campaign was conducted in 1994 and was comprised of observations carried out as part of two separate projects, AP263 and AP282 (PI: A.~Patnaik). Only four epochs were collected as part of AP282 which was predominantly used for conducting the JVAS observations. The second campaign (AH593, PI: J.~Hewitt) took place during 1996/1997 and targeted eight lens systems from which data on two other lens systems, JVAS~B0218+357 \citep{cohen00,biggs18} and CLASS~B1600+434 \citep{biggs21}, have been published to date. In all, 97 and 47 usable epochs were observed at 8.4 and 15~GHz respectively. The observations are summarised in Table~\ref{tab:obs}.

The VLA was used in standard continuum mode i.e.\ using two 50-MHz-wide spectral windows at an average frequency of 8.4 (8.5 during Season~2) and 15~GHz. The AP282 epochs were an exception as these used a narrower bandwidth of 25~MHz per spectral window. The correlator was configured to produce all four cross-correlations of the right and left circularly-polarized signals received by each antenna. Each monitoring season started in A~configuration and then moved into the smaller B~configuration, resulting in 8.4-GHz angular resolutions of $\sim$0.2 and 0.6~arcsec. The latter value is somewhat more than the angular separation of 0.5~arcsec between images A and B. The median 1422+231 observation time was $\sim$4 and $\sim$10~min during Seasons~1 (all frequencies) and 2 respectively.

The observing strategy was similar during both campaigns. As well as the lens itself, each epoch consisted of observations of a standard flux calibrator (3C~286) and a nearby ($\Delta\theta \sim 6.8$\degr) phase calibrator, OQ~208. The latter is compact \citep{stanghellini97} and, like many gigahertz-peak-spectrum (GPS) sources, demonstrates little flux variability \citep{odea98}. It is also unpolarized which allows it to be used as a polarization-leakage calibrator. The first season also included observations of 3C~287 which is not expected to vary as it is dominated by extended jet emission on scales of $\sim$100~mas \citep*{paragi98}. As such, it can be used to assess the quality of the flux calibration -- unfortunately no such source was observed during the second season. A notable difference with the AP283 epochs is that there were multiple (3--5) scans spread over a time interval of up to 7~hours. This allows the image variability on time-scales of a few hours to be probed, something that is particularly relevant in this system given the very short time delays predicted between images A, B and C.

The data were calibrated using NRAO's Astronomical Image Processing System (\textsc{aips}) following standard procedures. Briefly, gain solutions were found for all calibrators, with clean-component models used where these sources were resolved. 3C~287 is moderately resolved and we made our own model by combining data from multiple epochs and then mapping and self-calibrating in the usual way. The gain solutions were then used to find the flux density of all calibrators based on the assumed flux density of 3C~286 (using \textsc{getjy}) and the OQ~208 solutions interpolated onto the B1422+231 data. Polarization leakage (D-term) calibration was performed using \textsc{pcal} assuming that OQ~208 is unpolarized. The absolute angle of polarization was initially calibrated with \textsc{rldif} assuming the standard values for 3C~286. However, examination of the electric vector position angle (EVPA) of 3C~287 revealed the systematic hour-angle offset discussed by \citet{biggs18}. Therefore, we repeated the EVPA calibration for the first season's 8.4-GHz data (the 15-GHz SNR is too low for the effect to be significant) using 3C~287.

\begin{figure*}
  \begin{center}
    \includegraphics[width=0.33\linewidth]{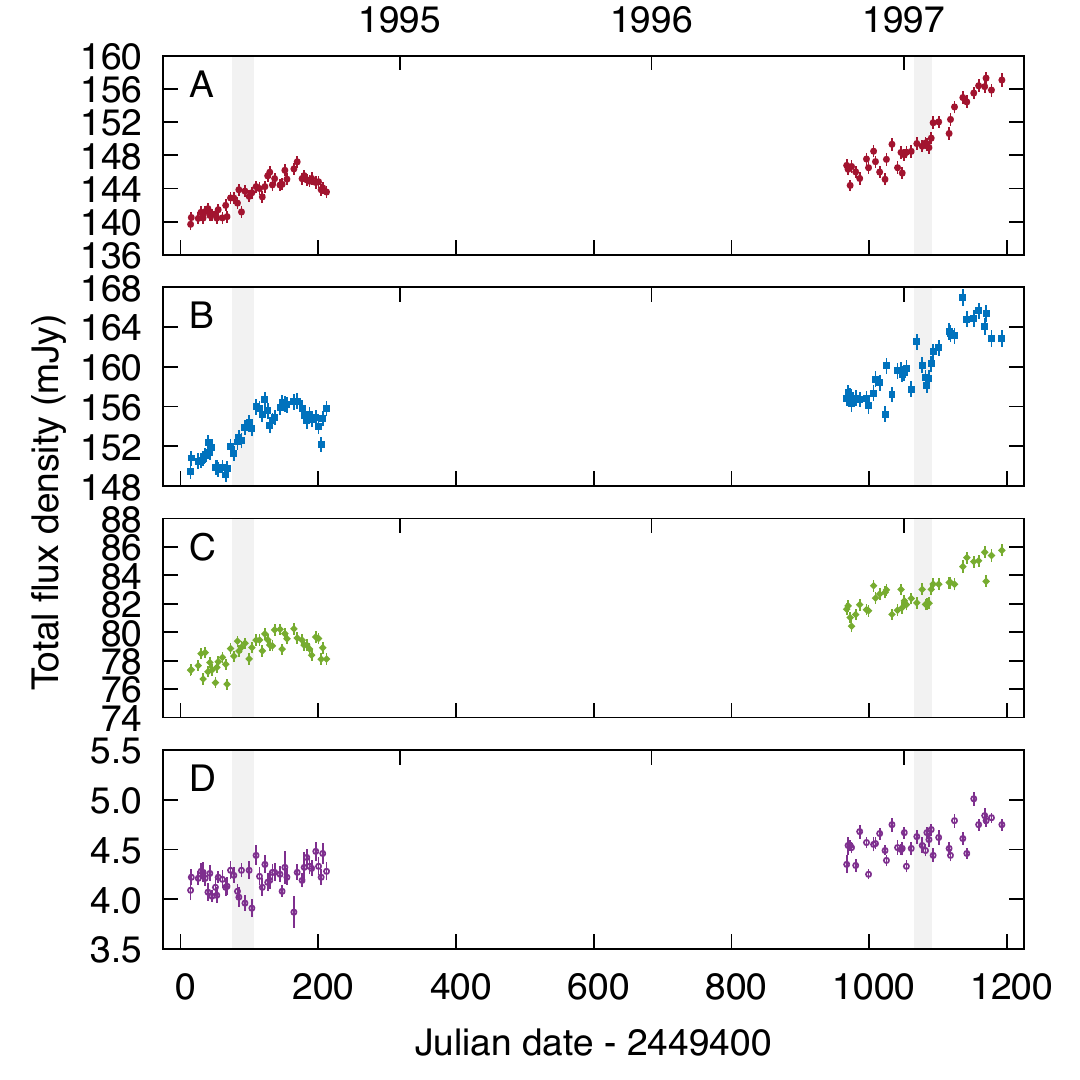}
    \includegraphics[width=0.33\linewidth]{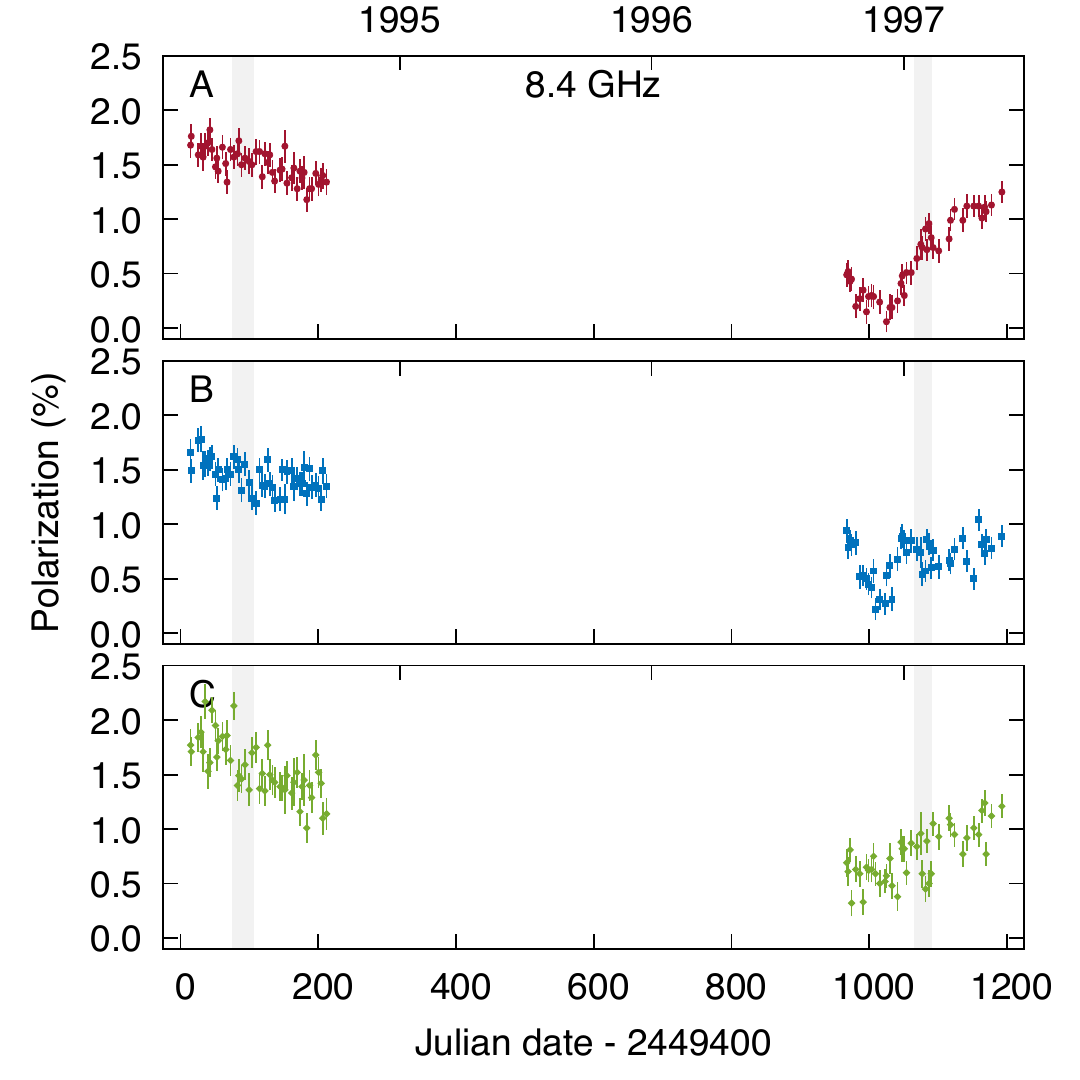}
    \includegraphics[width=0.33\linewidth]{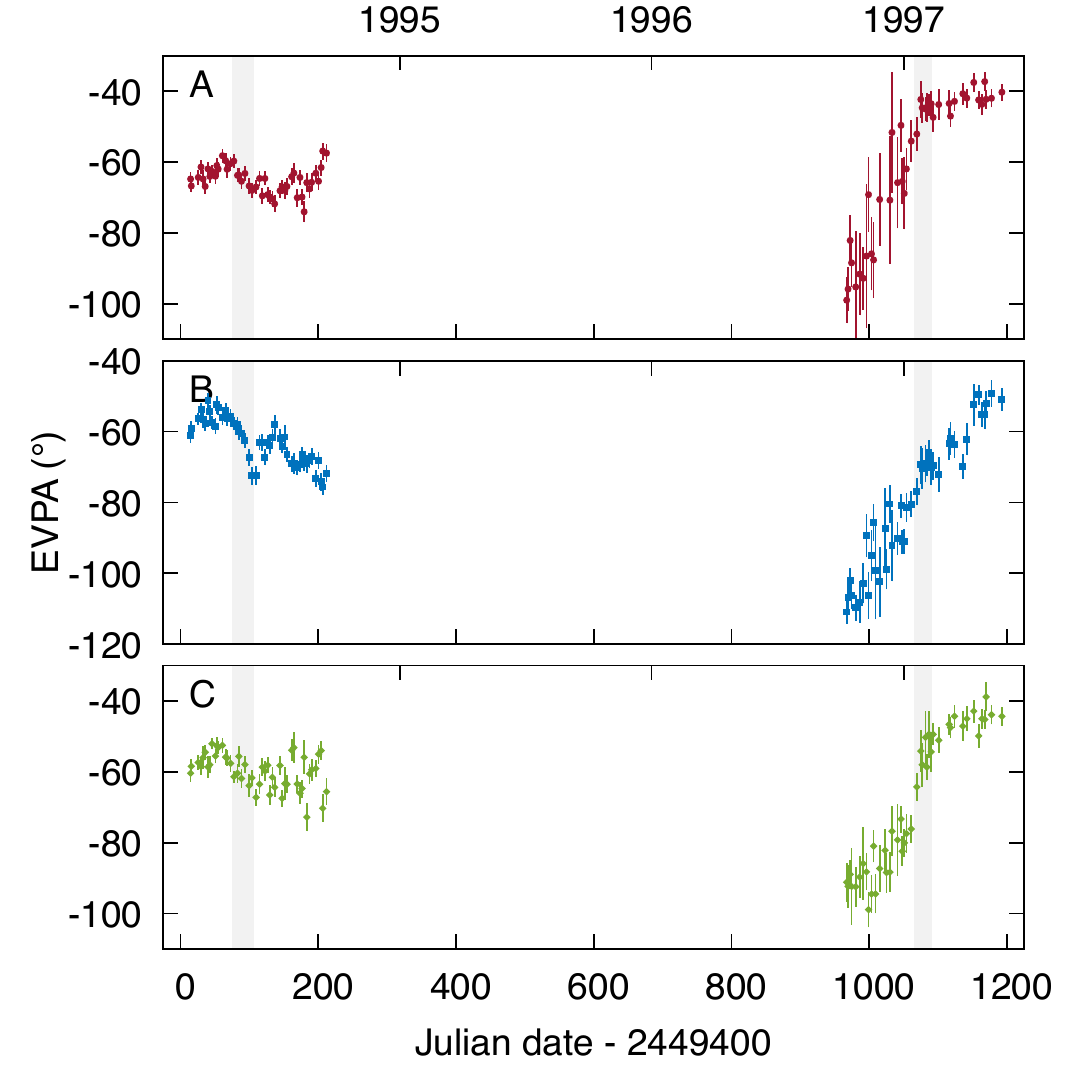}
    \includegraphics[width=0.33\linewidth]{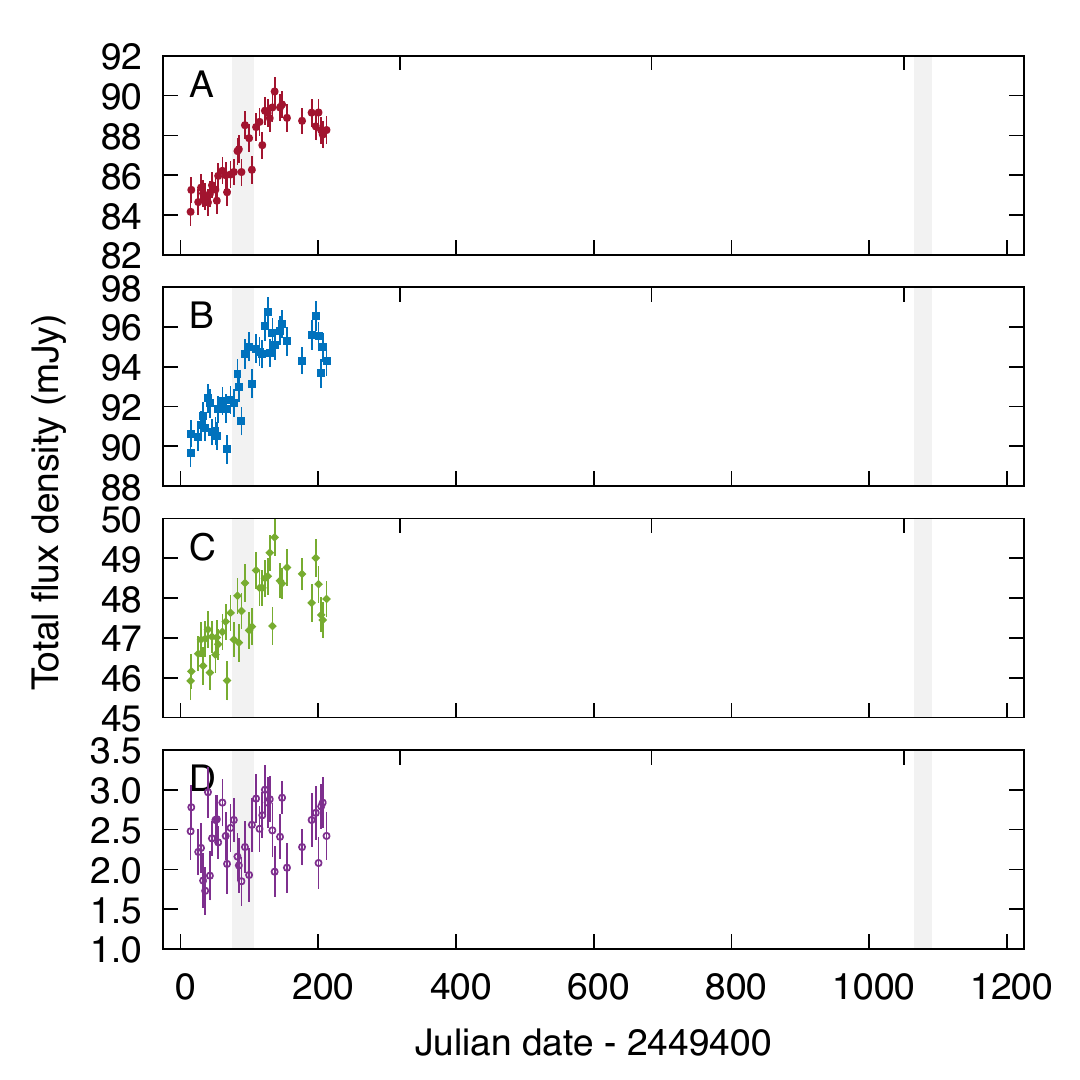}
    \includegraphics[width=0.33\linewidth]{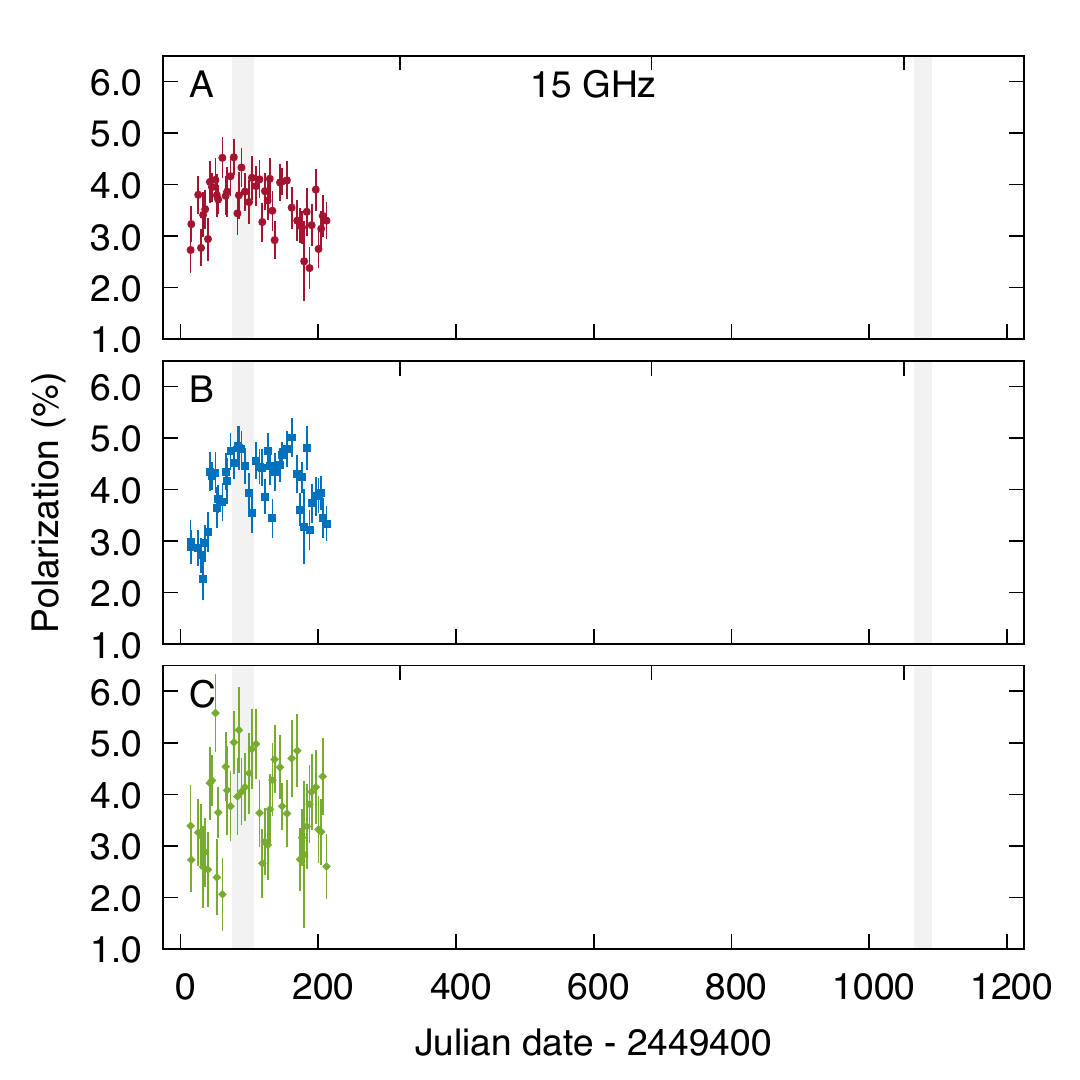}
    \includegraphics[width=0.33\linewidth]{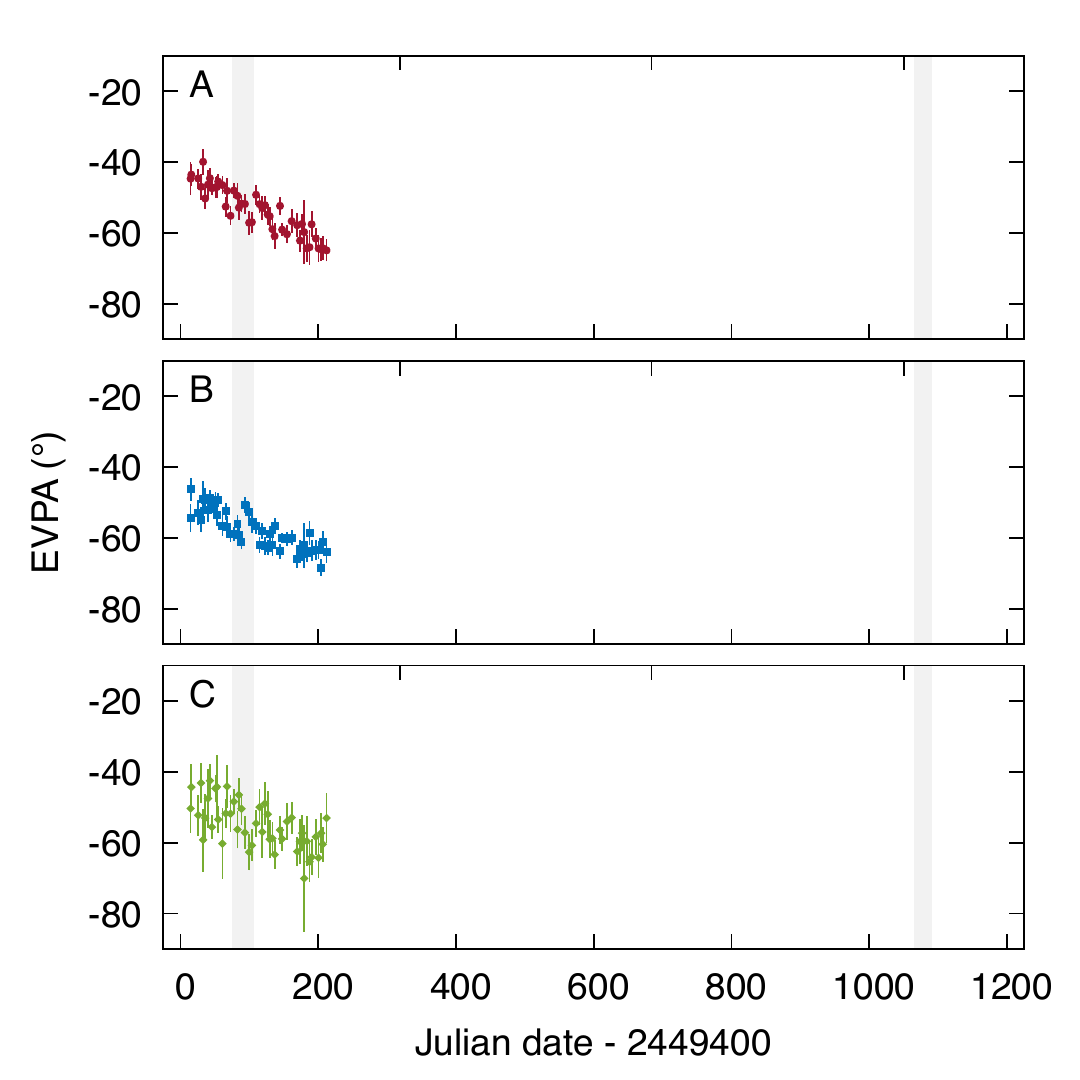}
    \caption{VLA radio light curves of B1422+231 at 8.4 (top) and 15~GHz (bottom). The polarization data for image~D are not plotted as the SNR per epoch is too low to allow a detection. Whereas the range of the \textit{y}-axis is dominated by the variability and the same for the three brighter images, for the total flux density of image~D this is dominated by the noise, especially at 15~GHz. The grey vertical bars indicate where the array was moving from A to B~configuration. The four epochs of multi-scan JVAS data at the beginning are not plotted.}
    \label{fig:vc_xu}
  \end{center}
\end{figure*}

\begin{figure*}
  \begin{center}
    \includegraphics[width=0.33\linewidth]{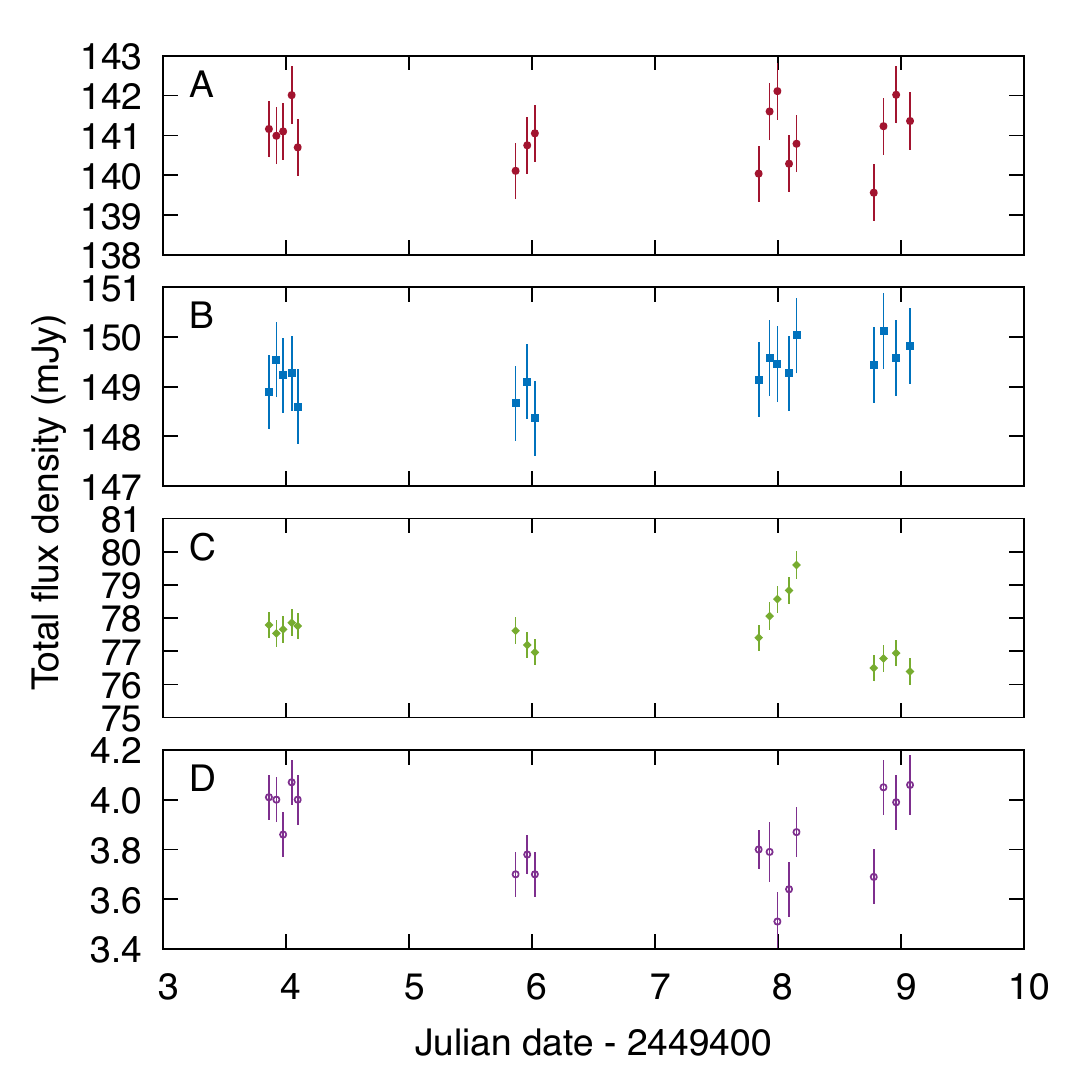}
    \includegraphics[width=0.33\linewidth]{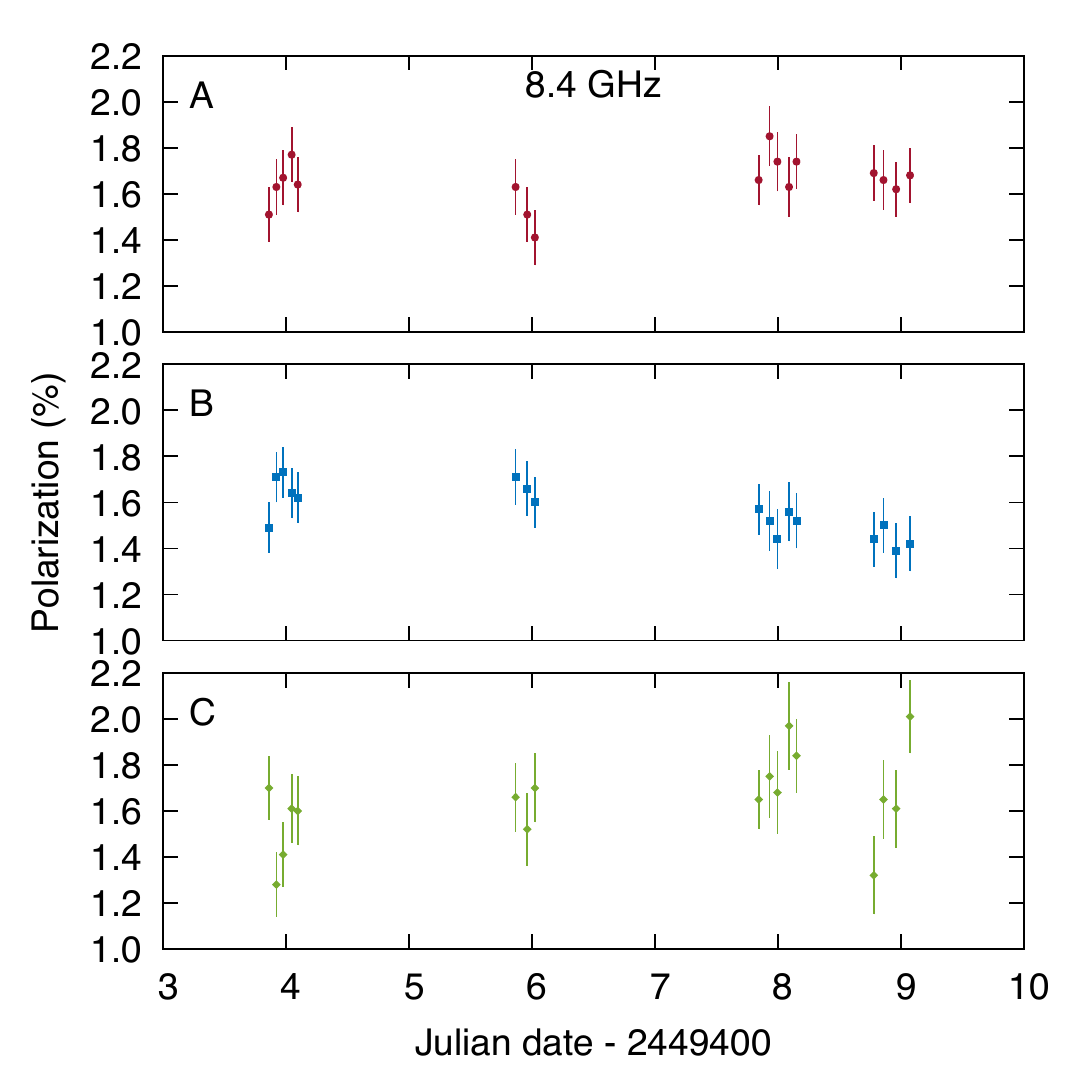}
    \includegraphics[width=0.33\linewidth]{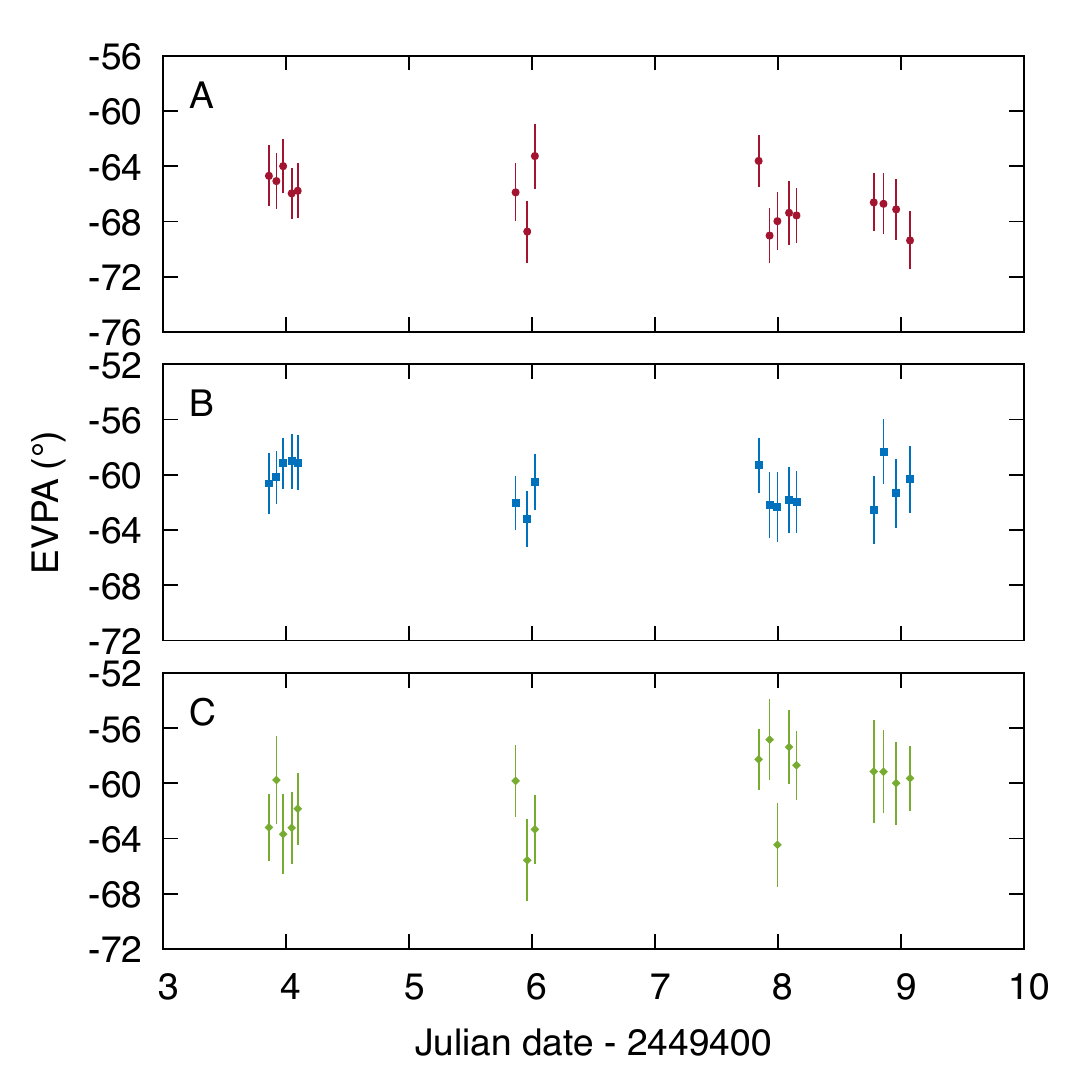}
    \includegraphics[width=0.33\linewidth]{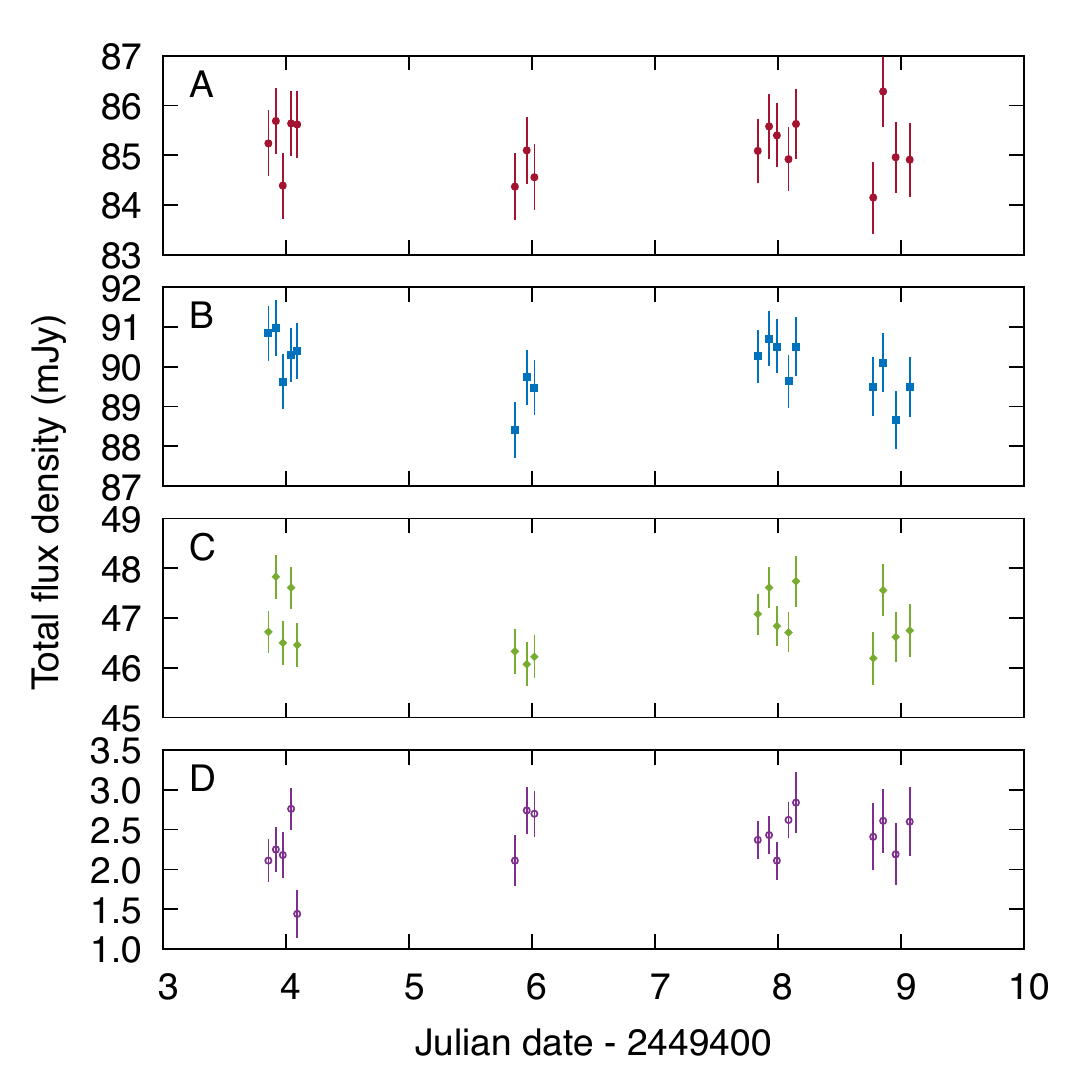}
    \includegraphics[width=0.33\linewidth]{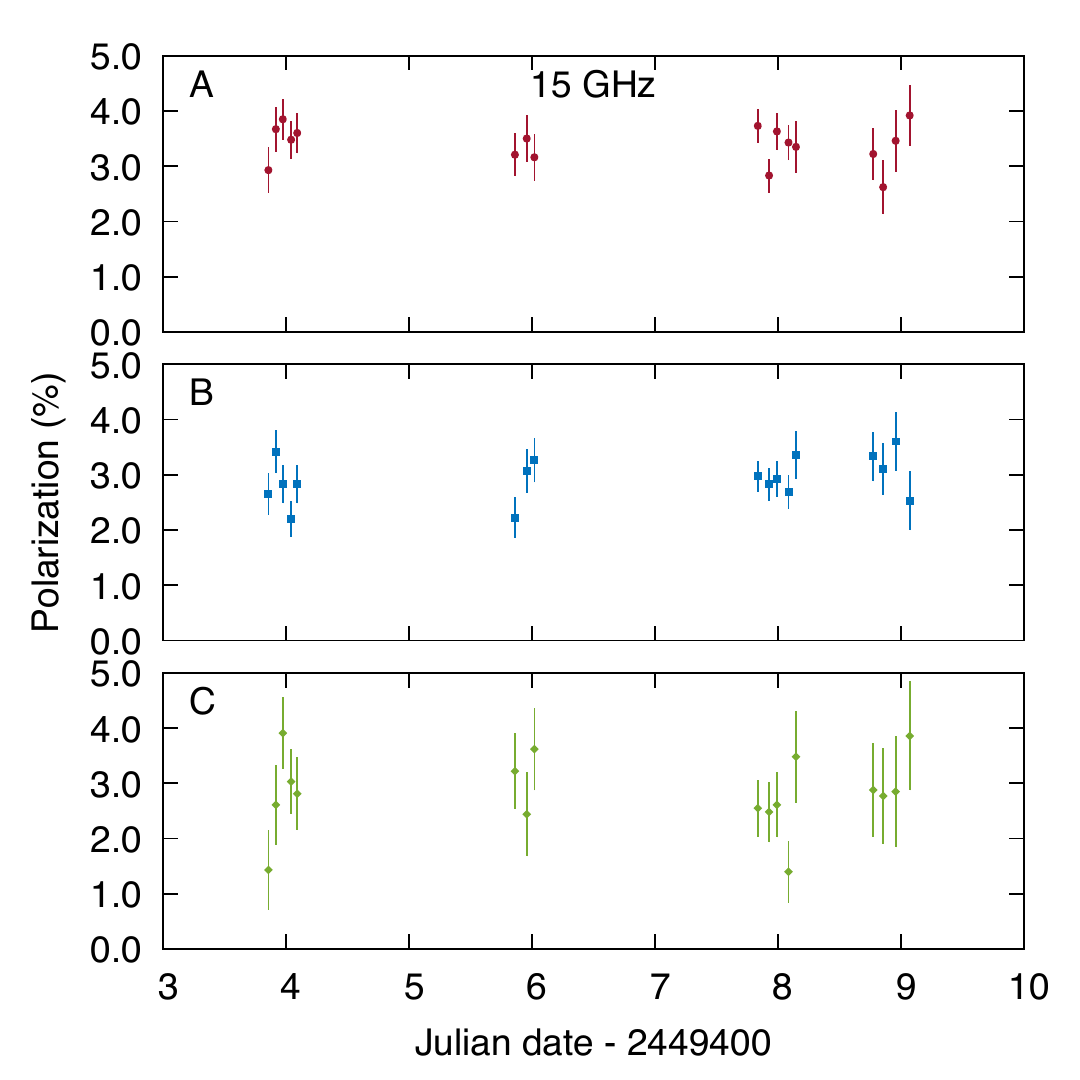}
    \includegraphics[width=0.33\linewidth]{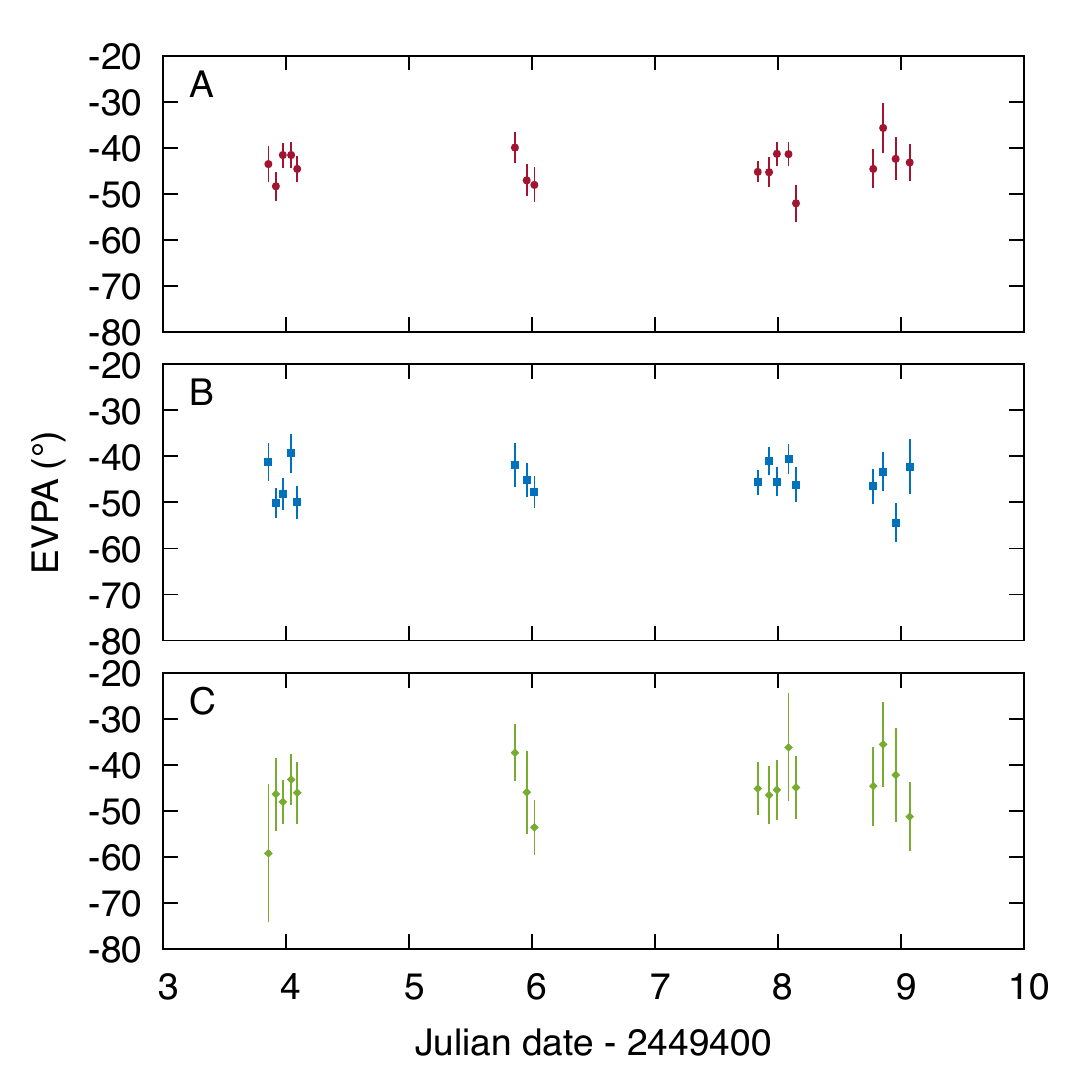}
    \caption{Radio light curves for the first four epochs that were observed as part of AP282 (JVAS). Top: 8.4~GHz, bottom: 15~GHz. These consisted of multiple scans and therefore allow intraday variability to be investigated. Most striking are the large and rapid variations of image~C in total flux density at 8.4~GHz. Image~D also varied by $\sim$10~per~cent over the course of the week.}
    \label{fig:vc_jvas}
  \end{center}
\end{figure*}

The flux densities of all four images were then calculated by modelfitting to the calibrated \textit{u,v} visibility data using Caltech's \textsc{difmap} package \citep{shepherd97}. As Very Long Baseline Interferometry (VLBI) has shown that each image has a very small angular extent at these frequencies \citep[$\la 5$~mas -- ][]{patnaik98,patnaik99}, each was modelled as an unresolved delta component fixed at its known position i.e.\ only the flux densities were allowed to vary. Modelfitting proceeded by alternating model optimization and phase-only self-calibration. This was repeated ten times at which point the phase corrections were removed and a final self-calibration performed using the last set of model parameters. Finally, we solved for a single amplitude correction per antenna and spectral window using the \textsc{gscale} command. This made very little difference to the flux densities of the bright components, but reduced the scatter in the much fainter image~D. The Stokes $Q$ and $U$ flux densities were then optimized using the self-calibrated data.

The radio light curves were then constructed from the raw $I$, $Q$ and $U$ flux densities with error bars calculated by combining an error in the flux density scale (0.5 and 0.7~per~cent at 8.4 and 15~GHz respectively) and the noise in a naturally weighted residual map in quadrature. Polarization errors were derived using the formulation of \citet{wardle74} and included a term representing residual polarization leakage, 0.1~per~cent of the Stokes $I$ flux density. Poor epochs were then identified by examining the rms noise, the reduced chi-squared of the fit to the visibility data and the flux densities of OQ~208 and, for the first monitoring season, 3C~287. We removed seven epochs at 8.4 and six at 15~GHz, all of the latter corresponding to a period of particularly bad weather towards the end of the first season. No polarization epochs were flagged, although for four epochs in Season~1 one of the spectral windows was flagged due to an instrumental problem that led to spuriously low polarization.

\section{Radio light curves}
\label{sec:rlc}

The 8.4 and 15-GHz radio light curves (without the JVAS epochs) are shown in Fig.~\ref{fig:vc_xu}. During the first season the total flux density of the three bright images increases by $\sim$5 and 7~per~cent at 8.4 and 15~GHz respectively, with a similar increase also occurring during Season~2 at 8.4~GHz. The most prominent deviation from this broadly linear increase is a small drop towards the end of Season~1 at both frequencies. This is the only obvious feature with which a delay might be measured, although it is not well sampled at 15~GHz due to flagging of epochs necessitated by bad weather.

Ideally, one would also wish to detect this peak in image~D, as the delay of about of a week predicted between it and the bright images should be easier to measure. However, the feature is not evident in D at either frequency. The 8.4-GHz data just show a gradual increase of $\sim$4~per~cent over the course of the season (so similar to that seen in the bright images before the peak), although the data are noisy -- the rms of the residuals around a straight-line fit is equal to approximately 3~per~cent of the average total flux density and would act to make the small decline at the end of the season difficult to detect. The 15-GHz data for image~D are even noisier ($5 < \mathrm{SNR} < 14$), but the data do not appear to be purely noise-like and there are signs that the total flux density is oscillating on a time-scale of about 50~days.

Greater intrinsic variability is seen in polarization. The polarized flux density of image~D is far too weak to be detected in a single epoch, but the other images are well detected at both frequencies -- at 8.4~GHz during Season~1 the images are detected with SNR$\ga10$. At this frequency, the magnitude of polarization gradually declines during the first season by about a fifth (the average value is $\sim$1.5~per~cent) and reaches a minimum near the beginning of Season~2 before beginning to increase again. At 15~GHz the source is more than twice as polarized, but the SNR is lower due to the higher noise. Despite this, there is a relatively sharp feature in the data towards the beginning of the monitoring when the polarization increases suddenly by about a third, most obviously in image~B. Significant variability is also seen in the EVPA, particularly in Season~2 where this rotates by $\sim$60\degr\ over a period of 6.5~months.

In addition to the intrinsic variability of the lensed source, there is also clear evidence that the images are varying independently of one another. As well as the oscillations at 15~GHz already discussed, in total flux density this is particularly pronounced in image~C during the first four (JVAS) epochs that consist of multiple short scans (Fig.~\ref{fig:vc_jvas}). The polarization also displays signs of independent variability, particularly during Season~2. This extrinsic variability is discussed in more detail in Section~\ref{sec:extrinsic}.

\subsection{Temporal smoothing of the \textit{u,v} data}

\begin{figure*}
  \begin{center}
    \includegraphics[width=0.31\linewidth]{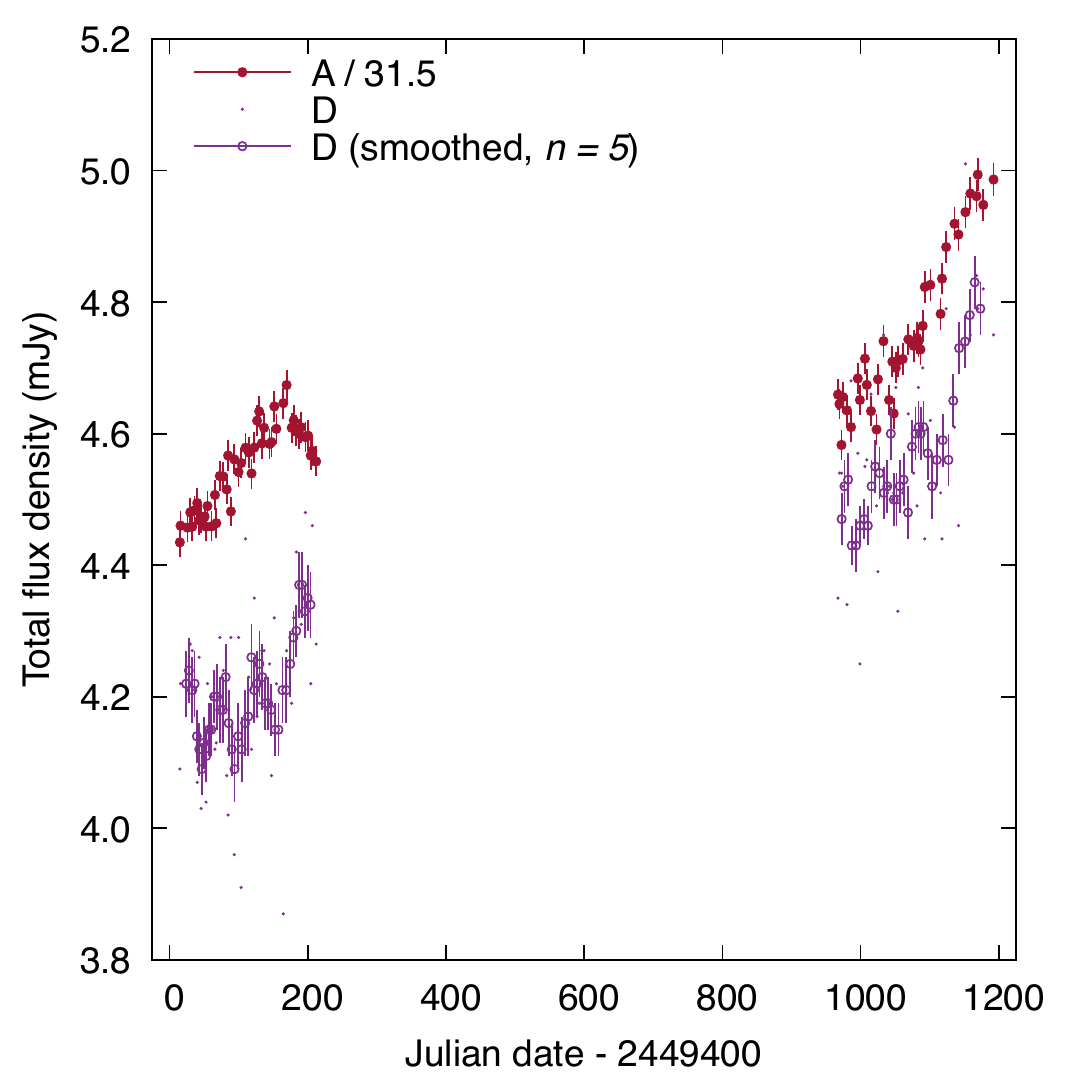}
    \includegraphics[width=0.31\linewidth]{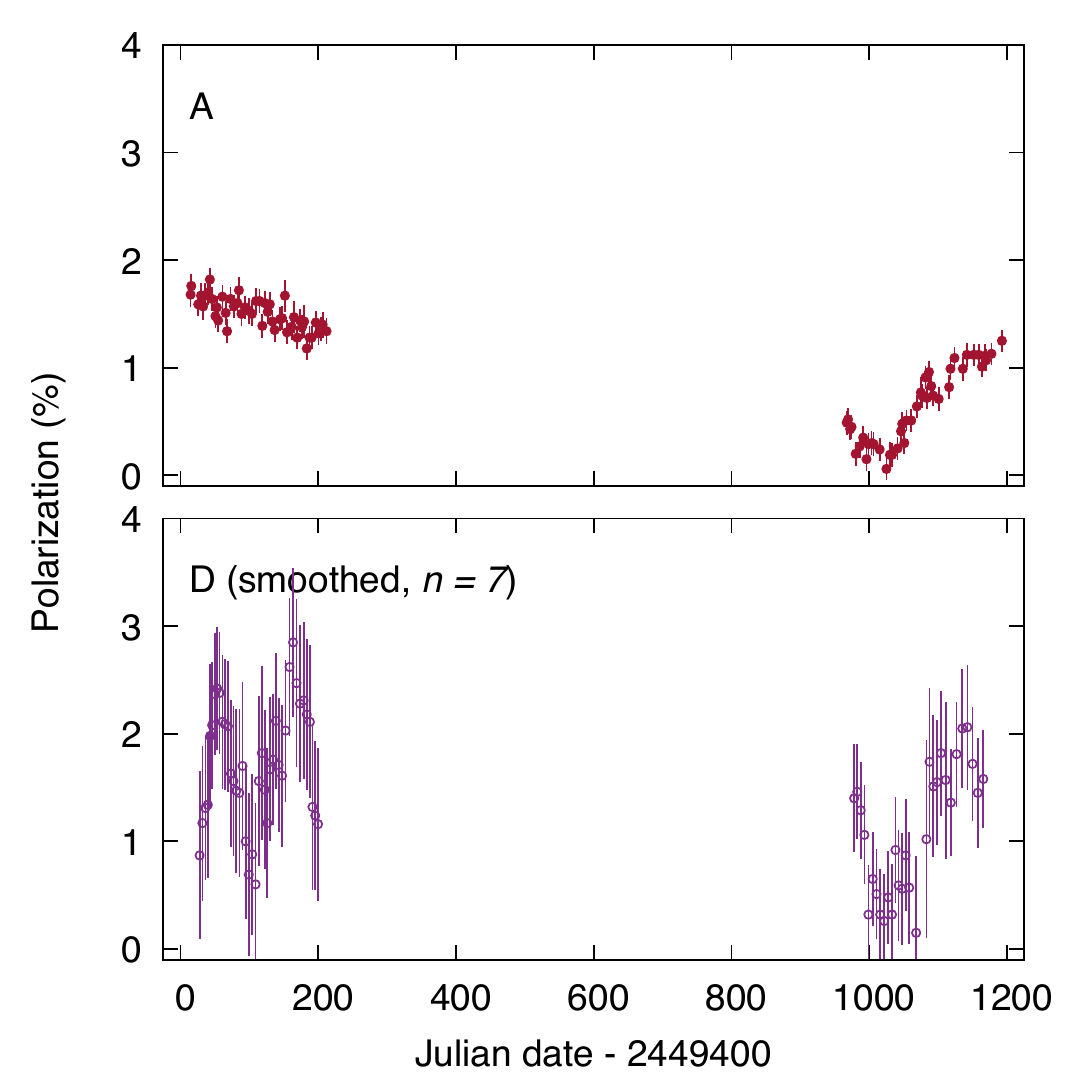}
    \includegraphics[width=0.31\linewidth]{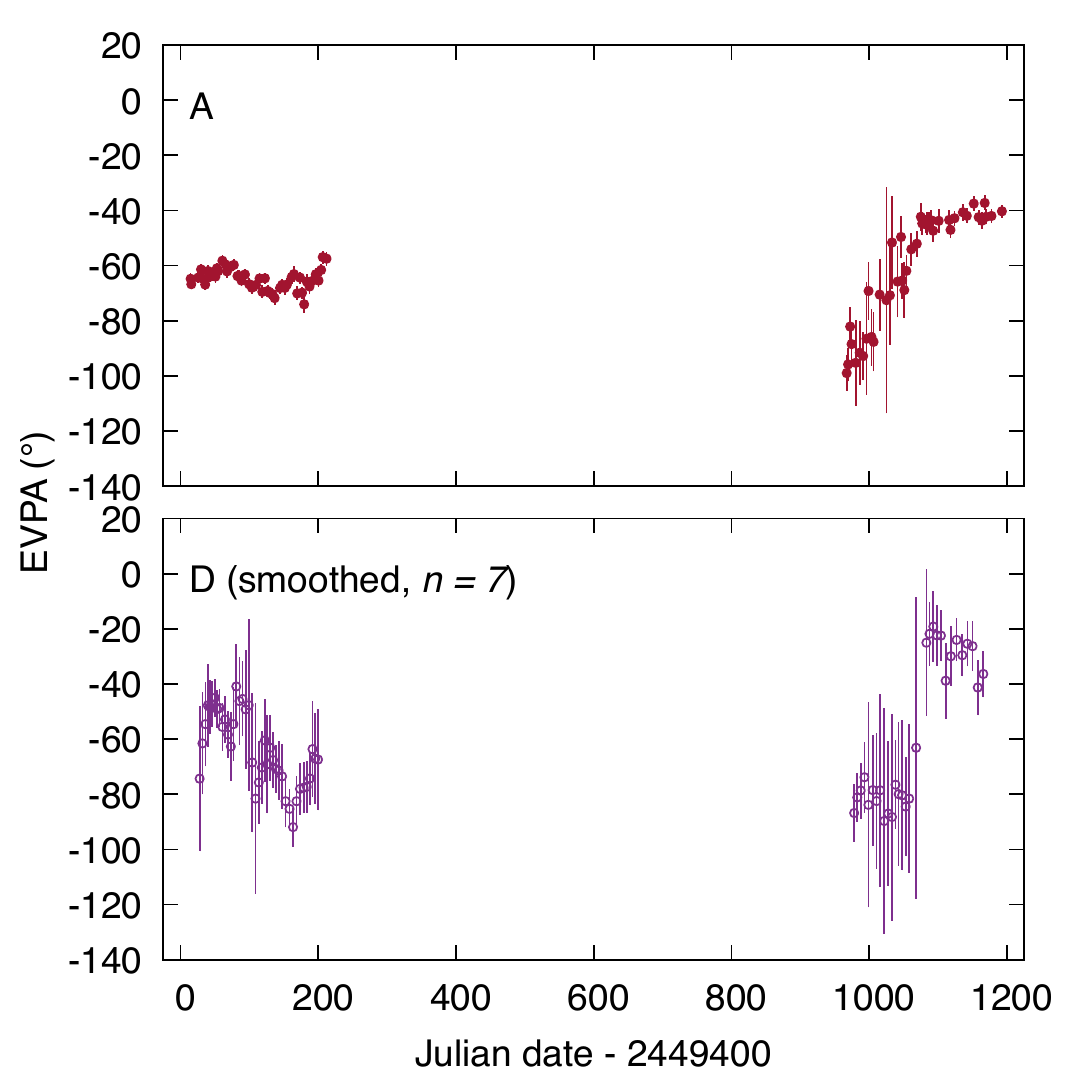}
    \caption{8.4-GHz data of image~D smoothed using a rolling average of $n$ epochs in the \textit{u,v} plane. From left to right: total flux density, percentage polarization and EVPA. The unaveraged data for image~A are shown by way of comparison and in the case of total flux density these have been divided by a factor of 31.5 -- note the change in the A/D flux density ratio with time. In Season~2, the averaging has revealed variations in D that are very similar to those seen in image~A. During Season~1, however, significant differences exist, especially in total flux density and percentage polarization. The total flux density plot also includes the unsmoothed image~D data, without errorbars for clarity.}
    \label{fig:vc_dsm}
  \end{center}
\end{figure*}

In order to increase the SNR of image~D, particularly in polarization, we have smoothed the 8.4-GHz \textit{u,v} data by forming new datasets each consisting of multiples of $n$ consecutive epochs. This has been done in the form of a rolling average whereby the first smoothed dataset consists of epochs 1 to $n$, the second of epochs 2 to $n+1$, etc. Before doing this, the fitted flux densities for images A, B and C are removed from each epoch's \textit{u,v} data using the {\sc aips} task {\sc uvmod} i.e.\ the smoothed datasets only include image~D. We use $n=5$ for total flux density and a larger value of $n=7$ for polarization due to the very low SNR. Once formed, the smoothed datasets are modelfitted in {\sc difmap} in the same way as the unsmoothed data.

The smoothed data are shown in Fig.~\ref{fig:vc_dsm} where we have also included the unsmoothed data of image~A as a comparison. After smoothing, the uncertainties on the total flux density measurements of images A and D are comparable and, during the second season, the variations in each image are roughly similar. During Season~1, though, image~D looks very different to A, being dominated by an apparent oscillation whereby it fades and brightens three times over the course of the season.

The smoothing has also improved the SNR of the polarization data, although the errors on image~D remain much larger than those on image~A. Despite this, we find variations in each image that are broadly similar during Season~2 with the positive rotation and the very low polarization close to the beginning of the season both detected. Season~1, on the other hand, again includes significant differences, this time in percentage polarization -- a prominent oscillation is seen in image~D, in contrast to image~A's slow, linear decline.

\section{Time-delay analysis}
\label{sec:delay}

The prospects for determining a time delay from the data presented in this paper do not appear promising due to a) limited intrinsic variability on short time-scales, b) the expected delays between the bright images being small compared to the average sampling interval between epochs, c) the faintness of image~D and d) the presence of extrinsic variability that produces independent variations in each image. It is interesting though to perform a time-delay analysis and compare these to the results of \citetalias{patnaik01b}, particularly as their measured delays were much larger than those predicted by lens modelling, at a significance level $>3 \sigma$.

We have investigated the time delay using the first season of total flux density data as this contains a clear feature that is present in all three bright images i.e.\ the peak close to the end of the monitoring. We exclude image~D from our time-delay analysis as the variations do not match those in the three bright images due to a combination of higher noise and, most obviously at 15~GHz, extrinsic variability. We do not include the JVAS epochs (AP282) as their sampling characteristics are very different to the rest of this season's data.

\begin{figure*}
\begin{center}
\includegraphics[width=0.8\linewidth]{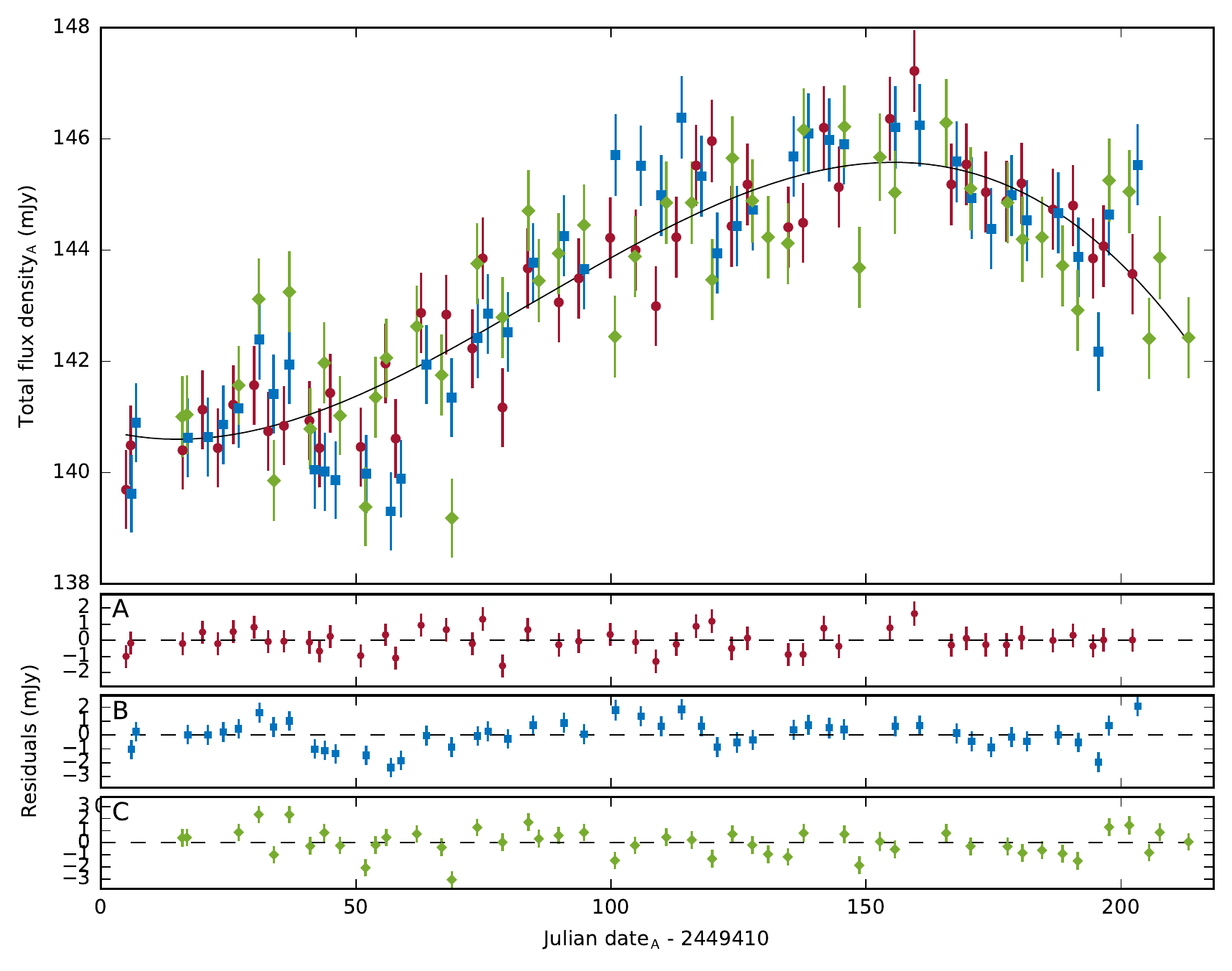}
\caption{Top: Total flux density of A, B and C after optimization of the time delays, flux ratios and coefficients of a cubic polynomial representing the intrinsic variability. Bottom: residuals around the optimized polynomial. Note the systematic variations in the residuals of B that occur between approximately days 40 and 120..}
\label{fig:tfx_poly}
\end{center}
\end{figure*}

\begin{table*}
  \centering
  \caption{Time delays ($\tau$) and flux ratios ($f$) relative to image~A and their uncertainties (1~$\sigma$) found using the total flux density data from Season~1. The Pelt method uses the $D^2_{4,2}$ statistic with $\delta = 5$~d.}
  \label{tab:tdelay}
  \begin{tabular}{cccccc} \\ \hline
    Method & Frequency (GHz) & $\tau_{\mathrm{B-A}}$ (d) & $\tau_{\mathrm{C-A}}$ (d) & $f_{\mathrm{B/A}}$ & $f_{\mathrm{C/A}}$ \\ \hline
    Poly & 8.4 & $-1.0^{+6.1}_{-5.3}$ & $-11.0^{+3.9}_{4.6}$ & $1.071 \pm 0.001$ & $0.549 \pm 0.001$ \\
    Pelt &     & $0.0^{+7.1}_{-7.4}$ & $-11.9^{+7.8}_{-8.7}$ & $1.071 \pm 0.001$ & $0.549 \pm 0.001$ \\
    Poly & 15  & $0.9^{+4.9}_{-4.4} $ & $-7.2^{+4.3}_{-4.5} $ & $1.072 \pm 0.002$ & $0.546 \pm 0.001$ \\
    Pelt &     & $0.0^{+6.6}_{-6.8}$ & $-1.1^{+7.8}_{-8.4}$ & $1.071 \pm 0.002$ & $0.546 \pm 0.001$ \\ \hline
  \end{tabular}
\end{table*}

The time delay has been calculated using two different methods. The first is inspired by the fact that the total flux density of image~A can be modelled using a simple cubic polynomial i.e. the chi-squared per degree of freedom ($\chi_{\mathrm{dof}}^2$) is equal to one. We therefore assume that the intrinsic variability can be represented by such a polynomial and optimize its coefficients together with the delays and flux ratios. In practice we first optimize the delays relative to image~A and then do the same for the flux ratios and the polynomial coefficients. This process proceeds iteratively until the delays have converged. This is similar to the `simultaneous spline fit' method of \citet*{tewes13}, as well as the spline fitting performed by \citetalias{patnaik01b}. However, the latter did not perform a `simultaneous' analysis and instead measured the three delays between each pair of images.

Our second method is the dispersion minimisation technique of \citet{pelt96} extended to systems with more than two images \citep{pelt98}. As with the polynomial method, we iteratively optimize the delays and flux ratios. We use the $D^2_{4,2}$ statistic \citep{pelt96} with a `decorrelation length' of $\delta = 5$~d, although the exact value makes little difference to the results. An oft-cited strength of the Pelt method is that it does not require interpolation in order to compare values in different images. This is particularly important when searching for delays that are expected to be close to zero due to the fact that interpolation is a smoothing operation. No interpolation is required at zero delay and the increasing amount of interpolation (and therefore smoothing) implied as the trial delay deviates from zero distorts the minimisation statistic.

For the error estimates we have produced simulated radio light curves by adding randomly generated residuals to a curve representing the intrinsic variability. As the Pelt technique is model-independent, we therefore fit a spline to the full dataset after removal of the best-fitting delays and flux ratios. This is forced to be rather smooth and in fact closely resembles the polynomial found by the first method. The residuals generally appear noise-like and thus the simulated residuals are formed by randomly selecting from a Gaussian distribution with a sigma equal to that of each epoch's uncertainty. For image~B at 8.4~GHz though, the residuals are highly non-Gaussian and thus we simulated residuals that conform to the statistical properties of the originals using the technique described in \citet{tewes13} which samples a time series produced using the method of \citet{timmer95}. We produce 1000 realisations of each light curve and use delays that are randomly perturbed within a $\pm$3~d window around the best-fitting delays.

Not included in the Pelt uncertainties is an additional term related to the semi-regular spacing of the epochs. These are often separated by integer multiples of 1~d and this characteristic spacing interacts with the finite width, $\delta$, of the time window within which neighbouring pairs contribute to the dispersion. The number of pairs contributing to the dispersion statistic tends to change abruptly at integer shifts of 1~d and this imprints local minima in the dispersion spectrum that are also separated by intervals of 1~d. This effect is discussed in more detail in \citet{biggs21} but with a magnitude of $\sim$0.5~d is small enough relative to the random uncertainties that its inclusion would make little difference to the final error estimate.

The results of the time-delay analysis are shown in Table~\ref{tab:tdelay} and the polynomial method illustrated in Fig.~\ref{fig:tfx_poly} where the 8.4-GHz data are shown after removal of the delays and flux ratios, together with the fitted polynomial and the residuals.

\section{Discussion}
\label{sec:discussion}

\begin{figure*}
  \begin{center}
    \includegraphics[scale=0.52]{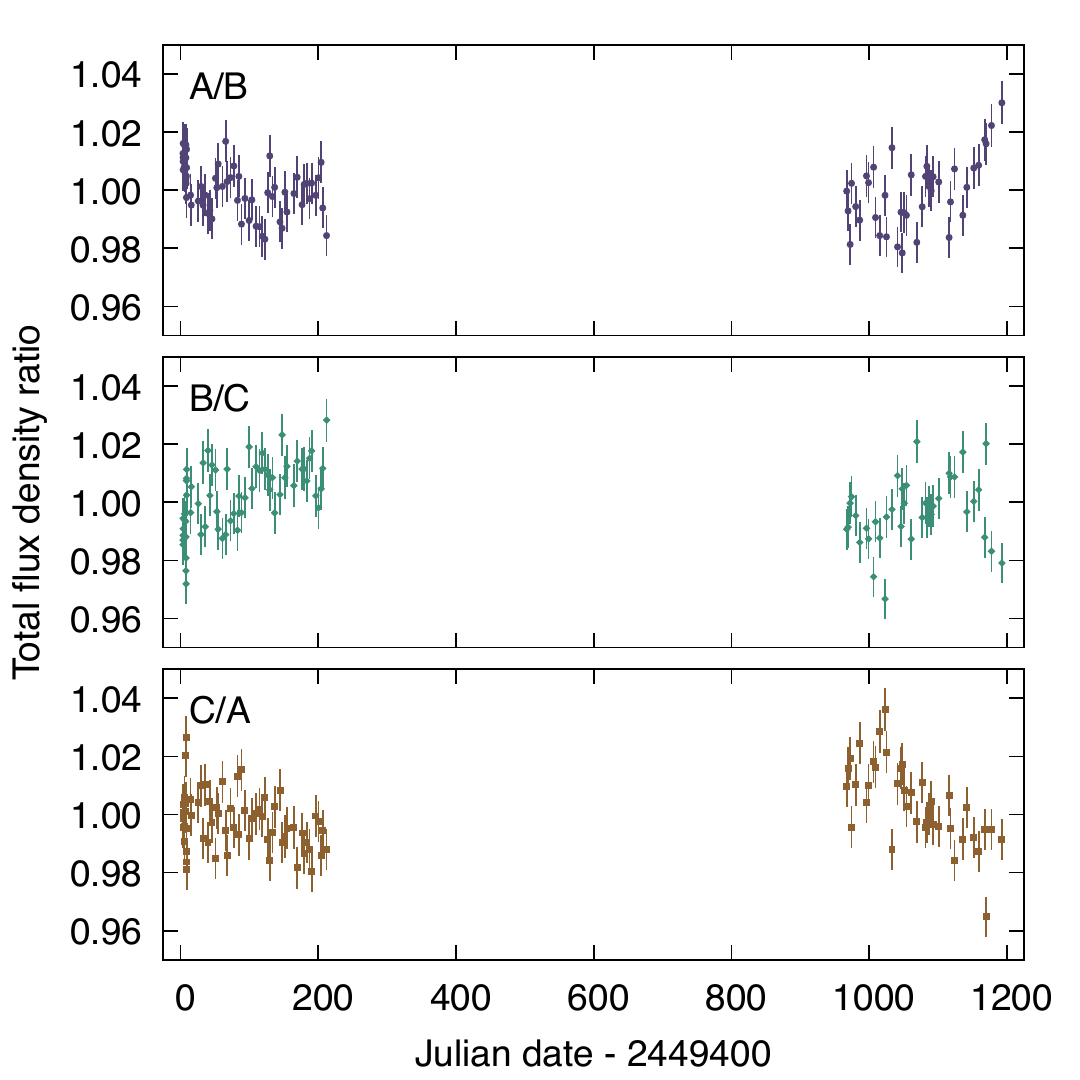}
    \includegraphics[scale=0.52]{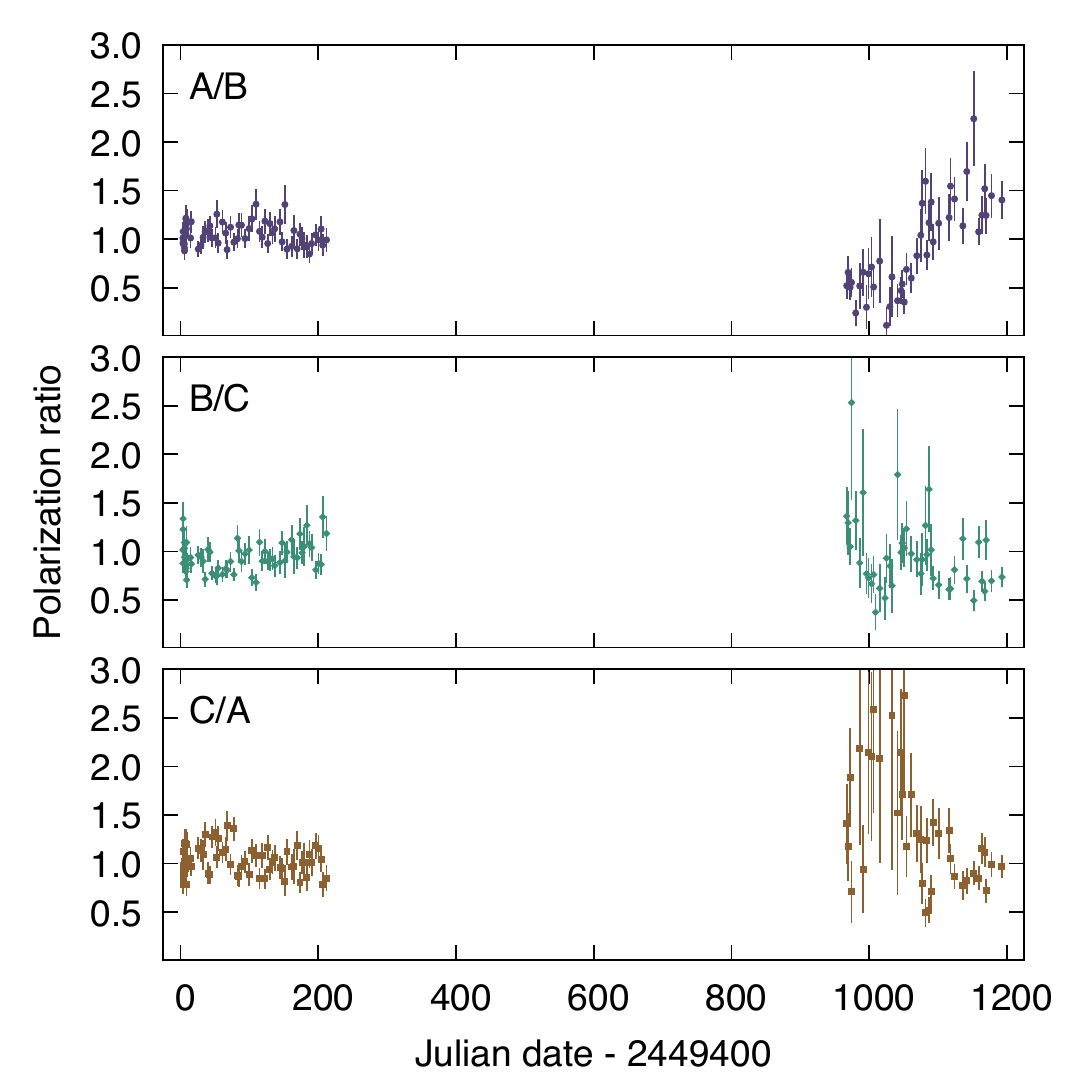}
    \includegraphics[scale=0.52]{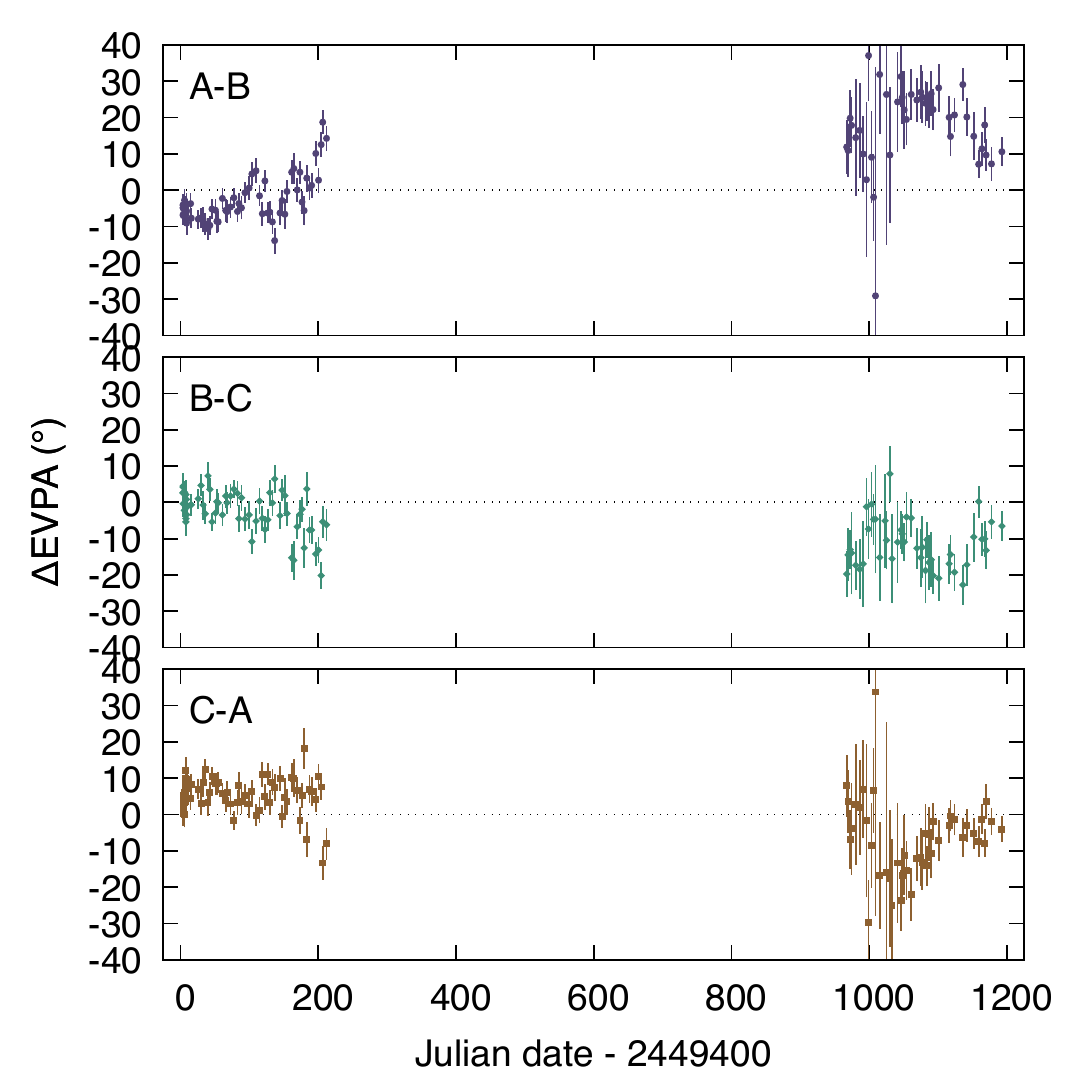}
    \caption{Ratio and difference plots for total flux density (left), percentage polarization (middle) and EVPA (right) at 8.4~GHz. The total flux density plots have been normalized by the time average of all points. With the intrinsic variability removed, the majority of the extrinsic variability during Season~1 can be seen to originate in image~B -- note the similar but mirrored variations in the A/B and B/C plots. Extrinsic variability is present in at least two images during the second season.}
    \label{fig:ratio_abc}
  \end{center}
\end{figure*}

\subsection{The time delay}
\label{sec:timedelay}

From the results presented in Table~\ref{tab:tdelay} we see that there is a certain amount of consistency between the results corresponding to different frequencies and methods. The delay between images A and B, $\tau_{\mathrm{B-A}}$, is essentially zero in all cases and whilst this is consistent with lens models, the errors are extremely large compared to the expected delay of much less than a day. The results for $\tau_{\mathrm{C-A}}$ are more interesting in that the data do seem to favour a large positive delay ($\ge 7$~d with C leading) in three out of four cases, similar to the results of \citetalias{patnaik01b}. However, the errors on our time delays are considerably larger and in all cases are consistent with the expected delay of $\sim 1$~d at a confidence level of $1-2~\sigma$. C leading A is, however, the predicted order and the fact that our data analysis and the independent one of \citetalias{patnaik01b} agree on this point is intriguing. 

Regarding the flux ratio, that between images A and B is the same at each frequency, as expected for a monochromatic process like gravitational lensing. For images A and C, however, the flux ratio C/A is lower at 15~GHz for each method. Whilst the significance is low ($\sim 1~\sigma$), it is possible therefore that some effect external to the lensed source is modifying the radio spectrum of one or more of the lensed images.

\subsection{Extrinsic variability}
\label{sec:extrinsic}

In presenting the radio variability curves and time-delay analysis, mention has been made of variability that is different in each image and which therefore is not intrinsic to the lensed source. We now go on to discuss this extrinsic variability in more detail.

We first examine the systematics in the time-delay residuals mentioned in Section~\ref{sec:delay}. Whereas the residuals shown in Fig.~\ref{fig:tfx_poly} for images~A and C are consistent with random noise, those of B show clear systematic trends. This image initially reduces in total flux density compared to A and C, before becoming relatively brighter about fifty days later. In total, the event lasts about eighty days. In a more statistical sense, the number of `runs' (sequences where the residuals do not change sign) is $\sim$4~$\sigma$ less than what would be expected for a random distribution \citep{wall03}.

This does not mean, however, that images~A and C are unaffected by extrinsic variability. Although the residuals for these images are normally distributed, the widths of these distributions are greater than would be expected if measurement error alone were the source of the scatter. Dividing each epoch's residual by its error bar should produce a set of points with a standard deviation of unity, but instead we find $\sigma = 1.5$ for both images. Image~D also contains more noise than expected, especially during Season~2 where the error bars are smaller due to longer observing times and here we measure $\sigma = 2.0$ around a straight-line fit. The reason for the extra `noise' could in fact be extrinsic variability occurring on a much shorter time-scale than the typical spacing between epochs -- the effect of the undersampling would be to add scatter to the data.

The question of whether very short time-scale variability is present can be investigated using the four multi-hour-angle JVAS epochs (Fig.~\ref{fig:vc_jvas}). At 8.4~GHz, the magnitude and time-scale of the variability in each image is very different. Image~B looks rather smooth and varies by $<$1~per~cent over the six days of monitoring. Image~C, on the other hand, shows remarkable systematic variations on time-scales of hours and this is most evident during the third epoch when its total flux density increases by 3~per~cent in seven hours. This event is not seen in the 15-GHz data, although this might be masked by the higher noise, $\sim$300 cf. $\sim$80~$\mu$Jy per epoch. Despite this, we can confirm that there is extrinsic variability taking place on very short time-scales in this system and which might be responsible for the larger than expected scatter in the monitoring data.

The extrinsic variability in total flux density and polarization for the three bright images during \textit{both} seasons is visualised in Fig.~\ref{fig:ratio_abc}. Here, we simply take the ratio or difference for the three pairs that can be formed from the image~A, B or C values. As the expected time delays are less than the spacing between epochs, the intrinsic variability should be removed in a model-independent way. Events that occur in a specific image will appear in two plots simultaneously, but as mirror images. The event visible in the time-delay residuals of image~B can be seen in both the A/B and B/C plots where the total flux density can be seen to oscillate with a magnitude of about 2~per~cent. During the second season, the most prominent example of extrinsic variability is again seen in image~B, at the very end of the monitoring. 

The polarization plots in Fig.~\ref{fig:ratio_abc} show more prominent extrinsic variations. There is little evidence of any independent variability during Season~1 in percentage polarization, but the EVPA of image~B gradually rotates by $\sim$25\degr, with smaller events superimposed. Season~2 also shows image-independent EVPA variability, but during this period the percentage polarization variations are also very different and for each this is happening in at least two images. Independent variability of each image's polarization properties is a natural consequence of extrinsic variations in total flux density if the intrinsic EVPA varies across the source and VLBI polarimetry has shown that this is indeed the case \citep{patnaik99}. However, there is little evidence of any correlation between the extrinsic variations in total flux density and polarization.

In order to put the assessment of extrinsic variability in total flux density on a more numeric footing, we turn to the method introduced by \citet{koopmans03}. This considers the vector, $r$, formed by assigning the normalized flux densities of A, B and C to the axes in a three-dimensional Cartesian coordinate system. As with the plots in Fig.~\ref{fig:ratio_abc}, intrinsic source variability (and other correlated variability such as that caused by errors in setting the flux scale) will not lead to a change in the direction of $r$, so long as the time delays are short compared to the spacing between epochs. If we define a plane that is normal to the (1, 1, 1) vector and project e.g.\ the A-axis onto this (naming this new axis $x$), independent variability in A will only move $r$ along the $x$-axis and in this way it is possible to isolate extrinsic variability occurring in a specific image. We refer the reader to \citet{koopmans03} for a full explanation of the method, the outputs of which are an estimation of the rms extrinsic variability of each image, as well as an indicator of the presence of extrinsic variability in the lens system overall i.e.\ not specific to a particular image.

The results for our VLA monitoring data are shown in Table~\ref{tab:extsig} where we include the 5-GHz results of \citet{koopmans03} and present results for the 8.4-GHz data as a whole, for the JVAS part and Seasons~1 and 2 separately. In all cases, the overall indicator of extrinsic variability, $\chi^2_r$, is greater than unity, thus supporting the visual impression that independent variability is present. This has its largest value for Season~2 at 8.4~GHz, $\chi^2_r = 5.9$, and an inspection of Fig.~\ref{fig:vc_xu} does show variations during this season that are less smooth than in Season~1. The only image for which the rms of the extrinsic variability is always greater than the rms of the data (and so can be measured) is image~C, the image where intraday variability is definitively detected at 8.4~GHz in the early JVAS epochs.

\begin{table}
  \centering
  \caption{Estimates of rms extrinsic variability in the three bright images, as well as an overall estimator ($\chi^2_r$) of whether extrinsic variability is present in \textit{any} image of B1422+231. Values are shown for all data at 8.4~GHz, as well as each season independently. Also included are the 5-GHz values derived by \citet{koopmans03} based on MERLIN monitoring. Values of $\chi^2_r > 1$ indicate the presence of extrinsic variability.}
  \begin{threeparttable}
    \label{tab:extsig}
    \begin{tabular}{ccccc} \\ \hline
      $\nu$ (GHz) & $\sigma_{\mathrm{A}}$ (\%) & $\sigma_{\mathrm{B}}$ (\%) & $\sigma_{\mathrm{C}}$ (\%) & $\chi^2_r$ \\ \hline
      5   & $-$\tnote{a}  & 0.6 & 0.9 & 3.7 \\
      8.4 & 0.5 & 0.5 & 0.8 & 4.8 \\
      8.4$_{\mathrm{JVAS}}$ & $-$\tnote{a} & $-$\tnote{a} & 0.9 & 3.3 \\
      8.4$_{\mathrm{S1}}$ & $-$\tnote{a} & 0.4 & 0.5 & 2.8 \\
      8.4$_{\mathrm{S2}}$ & 0.9 & 0.4 & 0.8 & 5.9 \\
      15  & $-$\tnote{a}  & 0.7 & 0.7 & 2.0 \\ \hline
    \end{tabular}
    \begin{tablenotes}
    \item [\textit{a}] These values are undefined as the expected rms is less than the rms of the data.
    \end{tablenotes}
  \end{threeparttable}
\end{table}

\begin{figure}
  \begin{center}
    \includegraphics[width=0.9\linewidth]{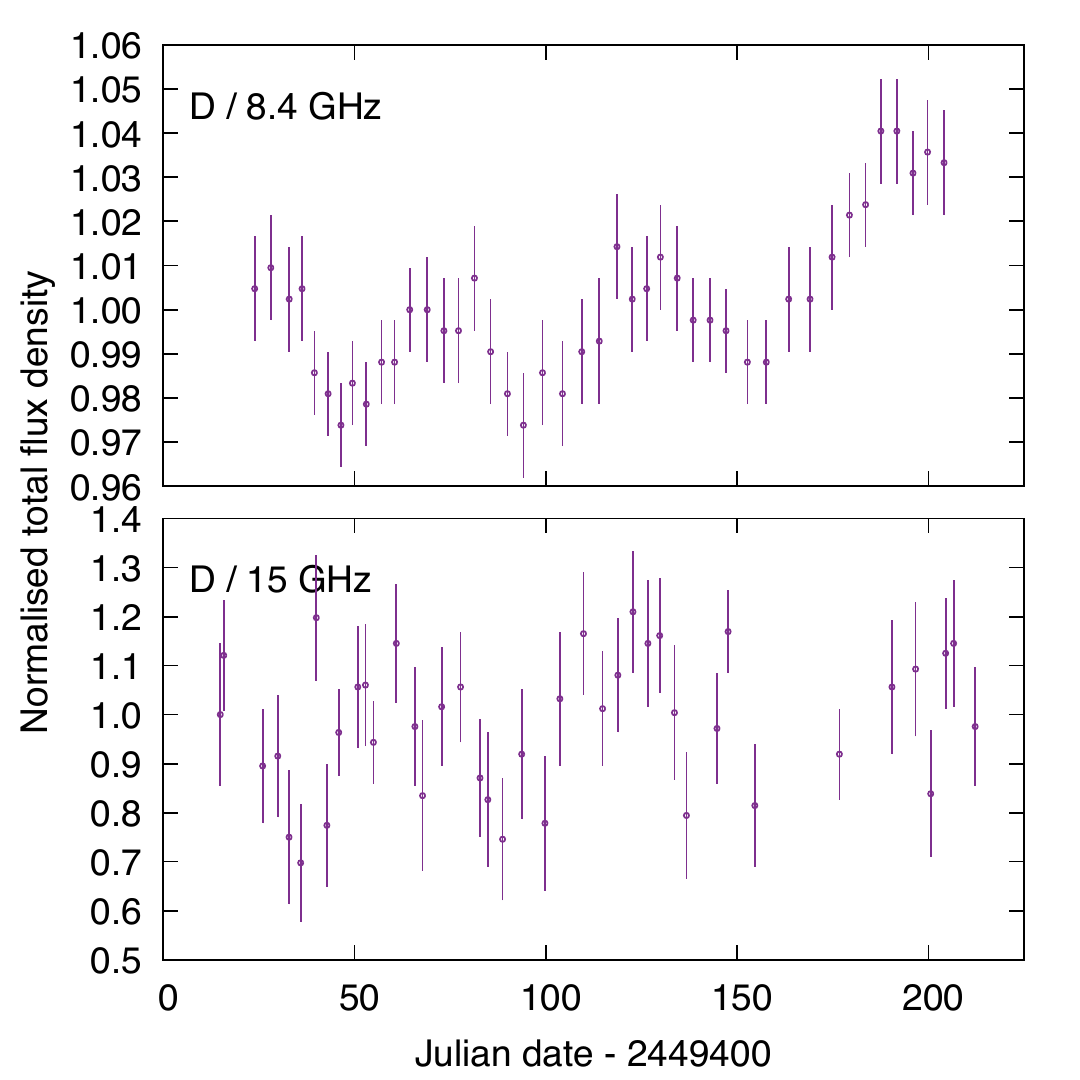}
    \caption{Total flux density of image~D at 8.4~GHz (top) and 15~GHz (bottom) with the data normalized by the average value in order to show the fractional variability. The 8.4-GHz data are the same as those shown in Fig.~\ref{fig:vc_dsm} and are smoothed using a rolling average of 5~epochs. An oscillation with a similar period seems to be present at both frequencies and with a magnitude that is much greater at 15~GHz.}
    \label{fig:d_oscillation}
  \end{center}
\end{figure}

Neither the plots in Fig.~\ref{fig:ratio_abc} nor the values in Table~\ref{tab:extsig} can reveal anything about extrinsic variability in image~D, as the larger time delay prevents a direct epoch-to-epoch comparison. However, we return to the observation in Section~\ref{sec:rlc} that the 15-GHz total flux density of this image varies in a different way to the brighter images, in that it appears to be oscillating. Given the large measurement errors relative to this image's total flux density, the oscillation must be quite significant and a plot of the normalized 15-GHz data for image~D normalized to its average total flux density is shown in Fig.~\ref{fig:d_oscillation}. This reveals that this image is varying by $\pm$10~per~cent around its average value with a time-scale (peak-to-peak) of about 50~days.

Also plotted in Fig.~\ref{fig:d_oscillation} are the \textit{smoothed} total flux density data for image~D at 8.4~GHz during Season~1. It has already been noted that these data also oscillate, but the comparison allowed by Fig.~\ref{fig:d_oscillation} strongly suggests that the two frequencies are brightening and fading with the same cadence, perhaps with 15~GHz leading the lower frequency by a small amount. The main difference though is the very much smaller amplitude of oscillation at 8.4~GHz, around $\pm$1~per~cent, an order of magnitude smaller than at the higher frequency.

Two generally accepted mechanisms for producing extrinsic variability in lensed images are scintillation in our galaxy or microlensing in the lensing galaxy \citep[e.g.][]{koopmans00b}. The former is caused by fluctuations in the density of free electrons along the line of sight to the lens system, with the velocity of the Earth relative to the scattering screen responsible for the time variability -- see e.g.\ \citet{narayan92}. Radio microlensing, on the other hand, is caused by the movement of compact components in a superluminal jet moving over the caustic pattern produced by a population of compact objects in the lens galaxy -- see e.g.\ \citet{gopal-krishna91}.

It is possible to distinguish between the two by establishing the frequency-dependence of the magnitude of the extrinsic variability \citep[e.g.][]{koopmans00b}. This is expected to increase with frequency for the case of microlensing as the smaller source size will lead to less averaging over a caustic pattern associated with the perturbing masses. For scintillation, the change is in the opposite sense and for the case of image~D therefore, the evidence appears to favour microlensing of a superluminal jet component associated with the lensed source.

\subsubsection{Microlensing}

{Exploring the microlensing option in more detail, we first note that the Einstein radius of a one-solar-mass point source given the source and lens redshifts and a standard $\Lambda$CDM cosmology\footnote{We assume $H_0 = 70$~km\,s$^{-1}$\,Mpc$^{-1}$, $\Omega_m = 0.3$ and $\Omega_\Lambda = 0.7$.} is $\theta_{\mathrm{E}} = 2.6~\mu$arcsec. In order for radio microlensing to be observed, the jet components would need to have a similar angular scale. The size of the components is determined by a number of factors including the fraction of the total source flux contained in the component ($f$) and the Doppler factor ($D$) of the jet. Taking the average value of these values found by \citet{koopmans00b} for the lens system CLASS~B1600+434 ($f = 0.07$ and $D = 2$) and using their equation~24, we find a component size of 2.4~$\mu$arcsec at 15~GHz, very similar to $\theta_{\mathrm{E}}$. So, whilst the details are somewhat uncertain, reasonable values for the parameters are compatible with microlensing being the cause of the extrinsic variations.

The largest magnitude of extrinsic variability is seen in image~D at 15~GHz with an observed oscillation of $\pm$10~per~cent. A similarly large oscillation would be easily seen in the brighter images and thus its absence suggests very different microlensing characteristics at the position of image~D. A potential reason for this is the much closer location of D to the centre of the lensing galaxy, which results in this image being seen through a higher surface mass density of both smooth and clumpy (stellar) matter.

In Table~\ref{tab:kappa} we show the values of the surface mass density (or convergence, $\kappa$) and the shear ($\gamma$) taken from the lens model of \citet{schechter14} for all four images. Also included is the stellar contribution to the surface mass density calculated by \citet{dogruel20} based on the results of \citet*{oguri14} who assess the relative contribution of the smooth dark matter and stellar components based on a statistical analysis of over a hundred strong gravitational lens systems. This gives a surface mass density in stars ($\kappa_{\ast}$) that is $\sim20-30$ times higher at the position of image~D, thus making this image more likely to be affected by microlensing. The details, however, are complicated \citep[see e.g.][]{dobler07} and detailed simulations would be required to fully investigate the microlensing in B1422+231.

\begin{table}
  \centering
  \caption{Properties of the B1422+231 lens model used in this paper. The convergence ($\kappa$) and shear ($\gamma$) are taken from \citet{schechter14} and the convergence due to clumpy stellar matter from \citet{dogruel20}, including the estimate of the distance of the image from the centre of the lensing galaxy (in units of the effective radius, $R_e$). Note the much higher stellar mass density ($\kappa_{\ast}$) at the position of image~D compared to the other images.}
  \label{tab:kappa}
  \begin{tabular}{cccccc} \\ \hline
    Image & $R / R_e$ & $\kappa$ & $\kappa_{\ast} / \kappa$ & $\kappa_{\ast}$ & $\gamma$ \\ \hline
    A & 3.239 & 0.380 & 0.098 & 0.037 & 0.473 \\
    B & 3.095 & 0.492 & 0.106 & 0.052 & 0.628 \\
    C & 3.382 & 0.365 & 0.090 & 0.033 & 0.378 \\
    D & 0.789 & 1.980 & 0.553 & 1.095 & 2.110 \\ \hline
  \end{tabular}
\end{table}

\subsubsection{Scintillation}

Scintillation does not appear to be a likely candidate to explain the extrinsic variability, but in order to add weight to this conclusion, we have used the model of \citet{cordes02} that seeks to describe the fluctuations in the distribution of free electrons in the Galaxy. B1422+231 has a high Galactic latitude, $b = +68\fdg5$, and its coordinates correspond to a correspondingly low expected transition frequency between strong and weak scattering of $\nu_0 = 6.8$~GHz. Therefore, all data presented in this paper should be located in the regime of weak scattering. Assuming that this is indeed the case, the equations given in \citet{walker98,walker01} predict a modulation index (rms fractional flux variation) for a point source at 15~GHz of $m_p = 0.33$ and a time-scale for the variability, $t_p$, of 1.3~hours. Whilst the modulation index is only a factor of two larger than the observed value for the oscillation seen in image~D (15~per~cent) the discrepancy in the time-scale is $\sim 10^3$ and as such it is impossible to get both $m_p$ and $t_p$ to agree with the model predictions by e.g.\ varying the size of the varying component.

Even though microlensing seems the most likely explanation, we note that the picture of extrinsic variability in this system is complicated i.e.\ there appear to be independent variations affecting different images with different time-scales and amplitudes, properties which themselves appear to vary with time. More extensive multi-frequency monitoring is required before a complete picture of the extrinsic variability in this system will emerge.

\subsection{A radio arc between images B and C}

\begin{figure*}
  \begin{center}
    \includegraphics[scale=0.35]{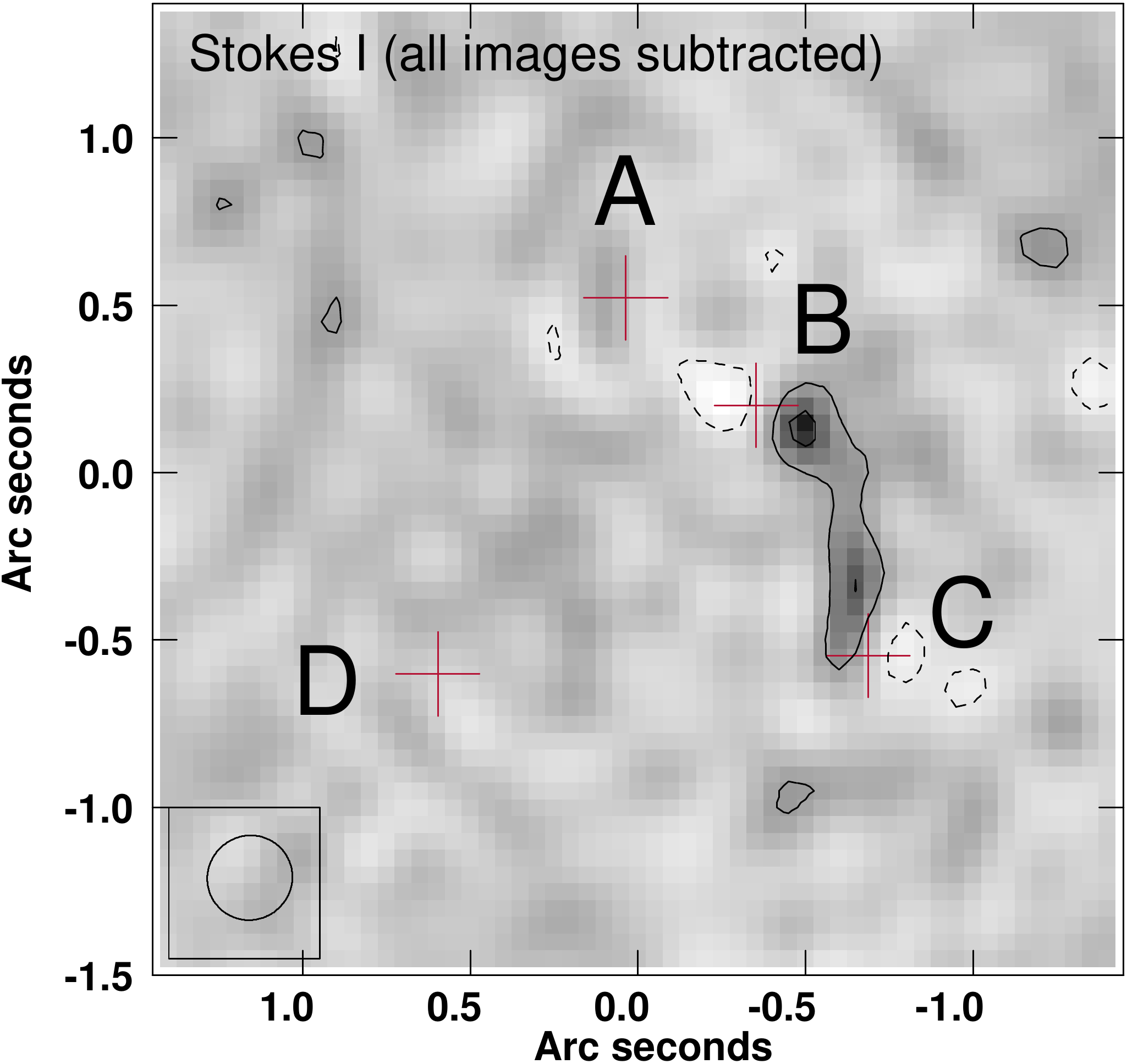}
    \includegraphics[scale=0.35]{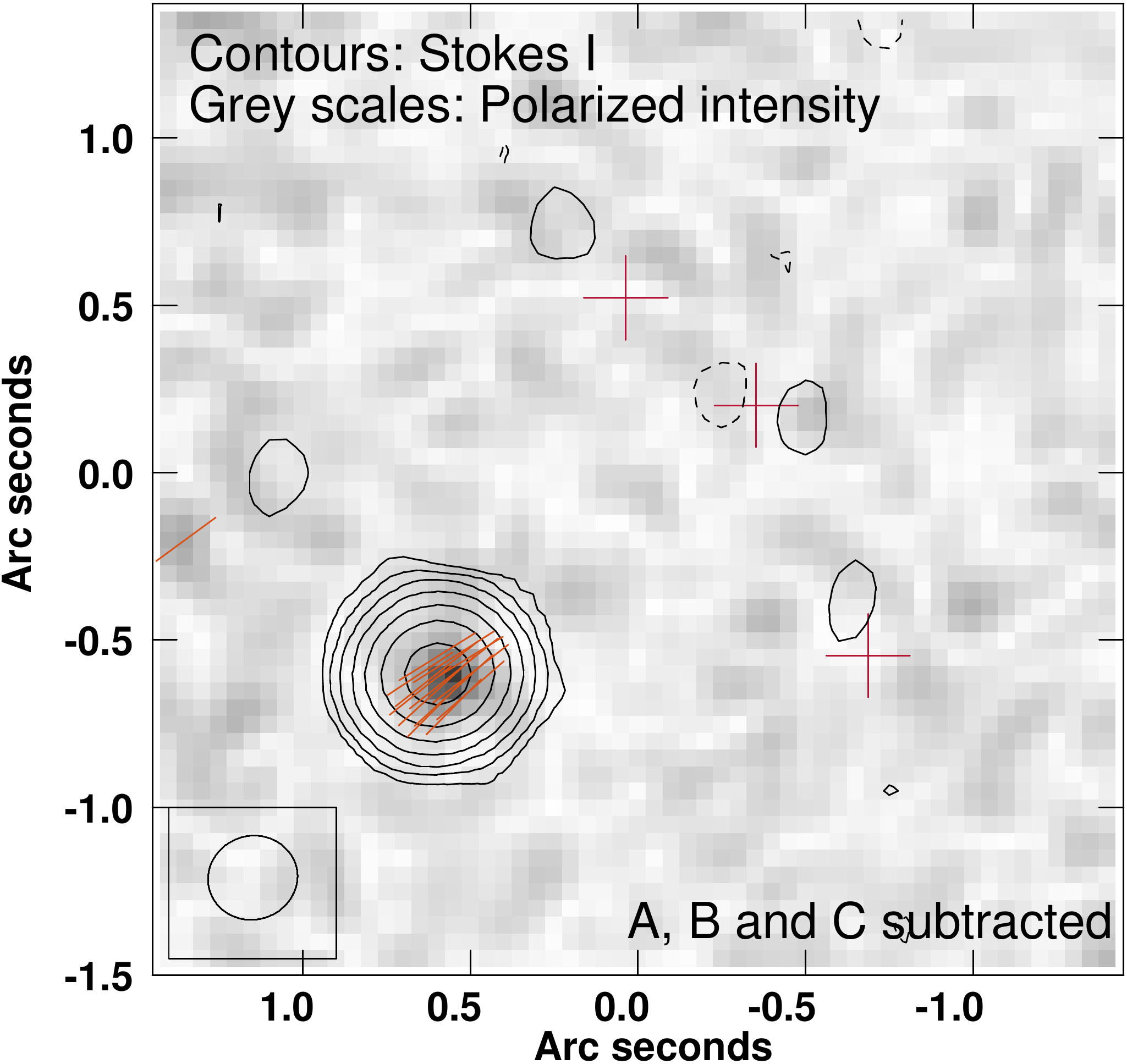}
    \caption{Left: Map made from combining all 8.4-GHz epochs after first subtracting the lensed images. Contours and grey scales both show Stokes $I$ and the extremely low noise (9.4~$\mu$Jy\,beam$^{-1}$) reveals a faint arc linking B and C. Right: Map made from combining all 8.4-GHz epochs from Season~1 after subtraction of only the three brightest images. Contours show Stokes $I$ and grey scales the polarized flux. Pixels where the polarization is detected at 3~$\sigma$ or higher are shown with sticks and demonstrate that polarization is detected in image~D. Image positions are marked with crosses. The restoring beam of each image is $\sim0.26 \times 0.25$~arcsec$^2$ and contours are plotted at $-1$, 1, 2, 4. etc. multiples of three times the rms noise.}
    \label{fig:stuffr}
  \end{center}
\end{figure*}

\begin{figure*}
  \begin{center}
    \includegraphics[scale=0.21]{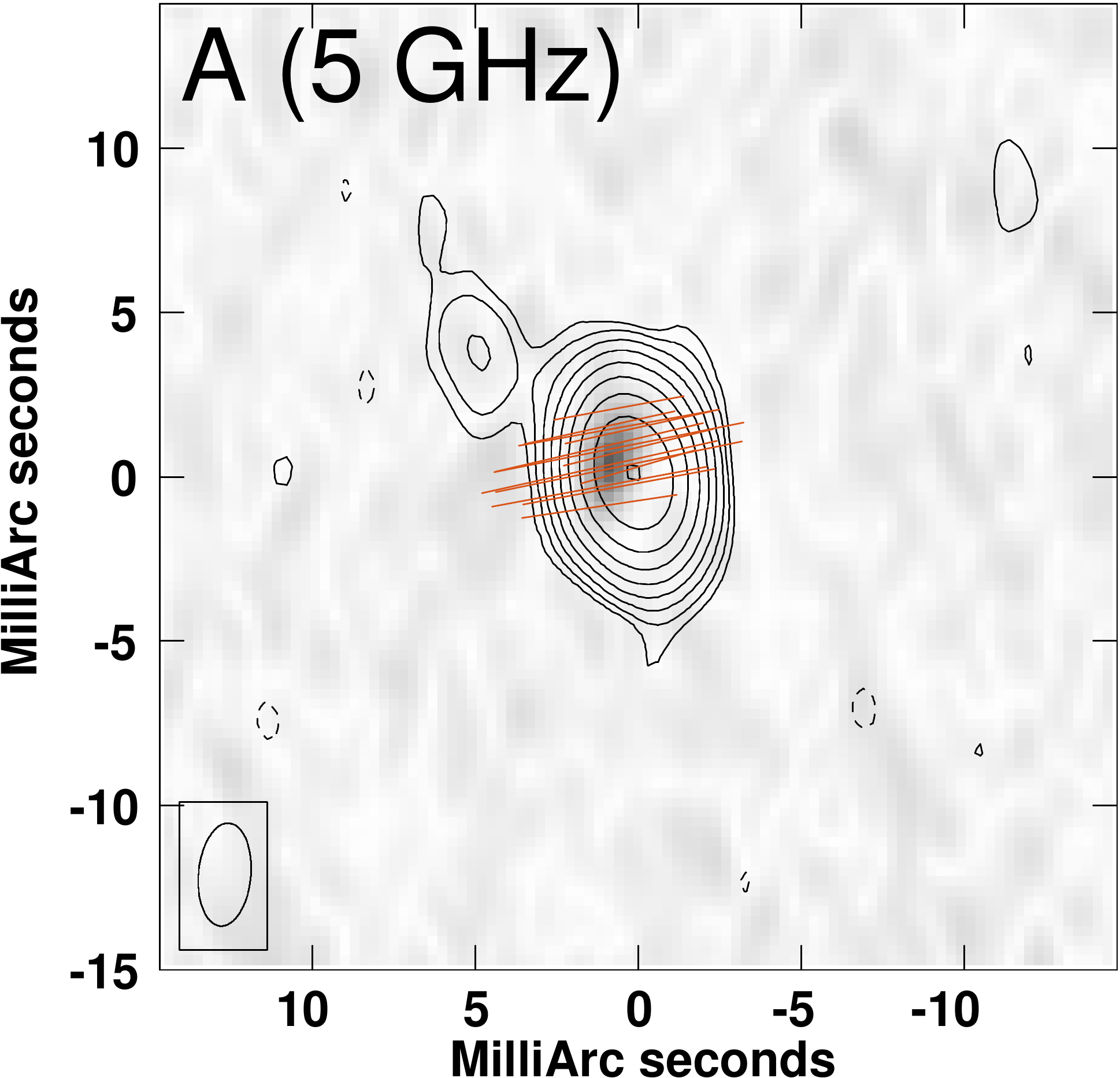}
    \includegraphics[scale=0.21]{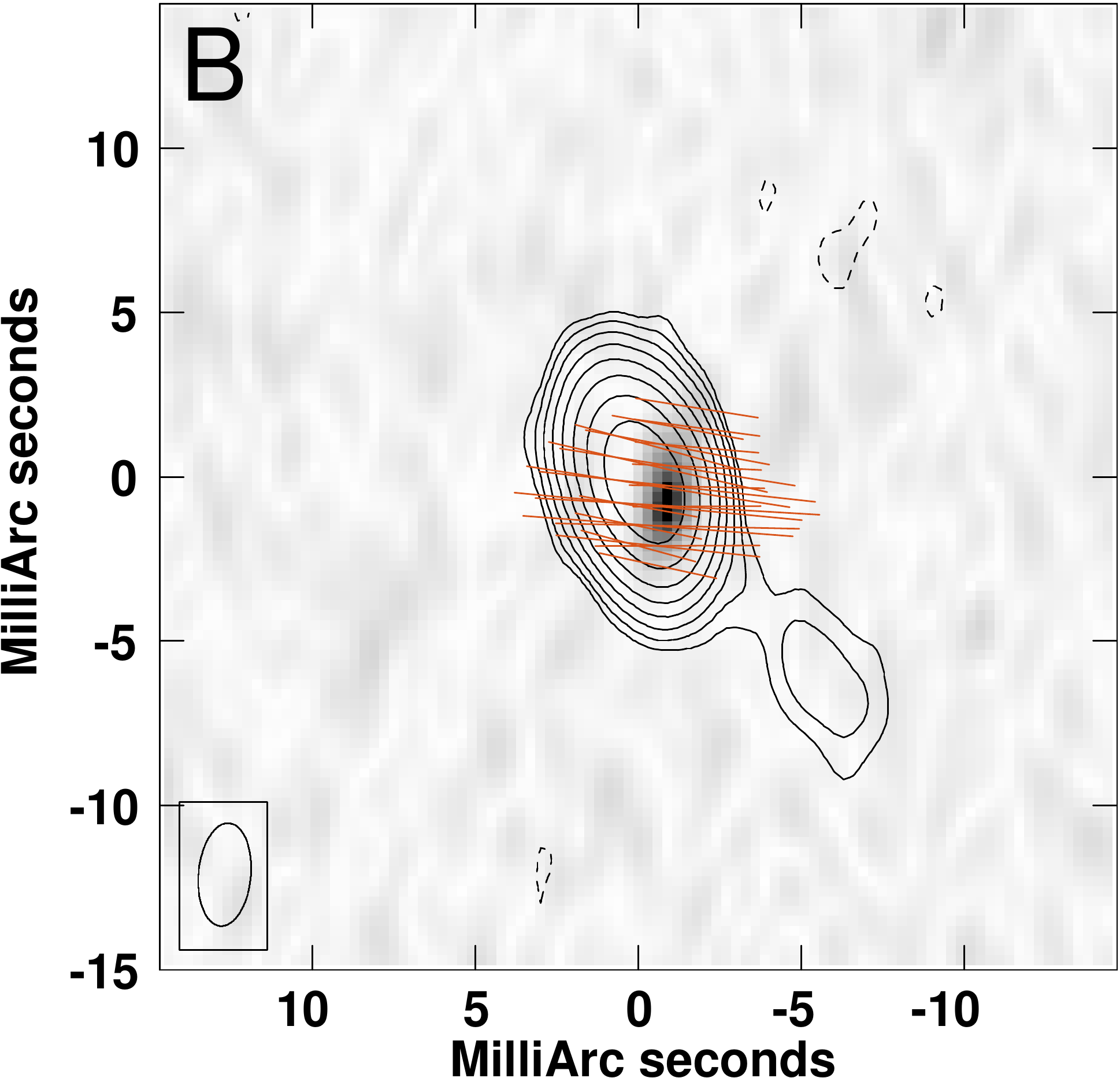}
    \includegraphics[scale=0.21]{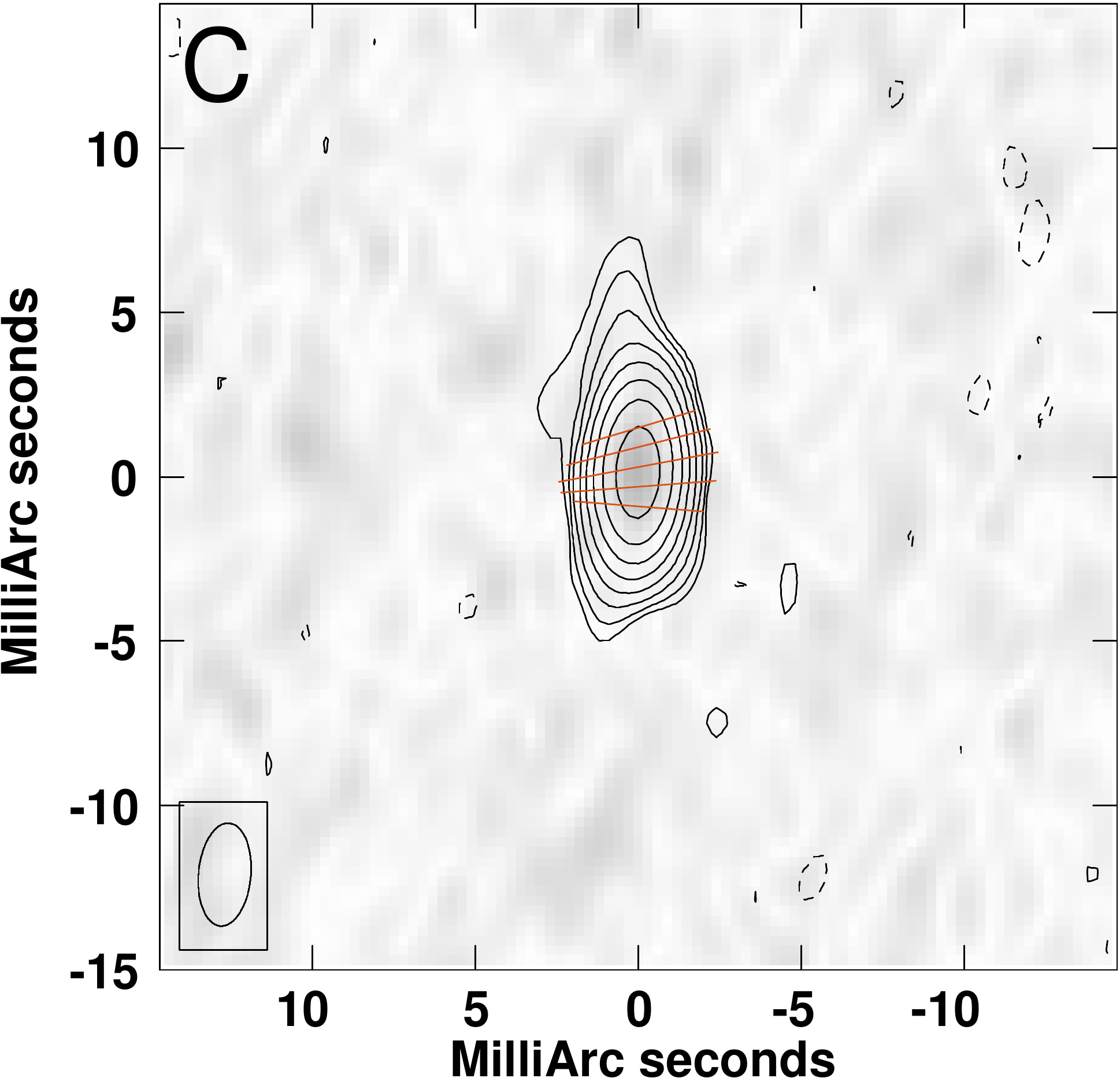}
    \includegraphics[scale=0.21]{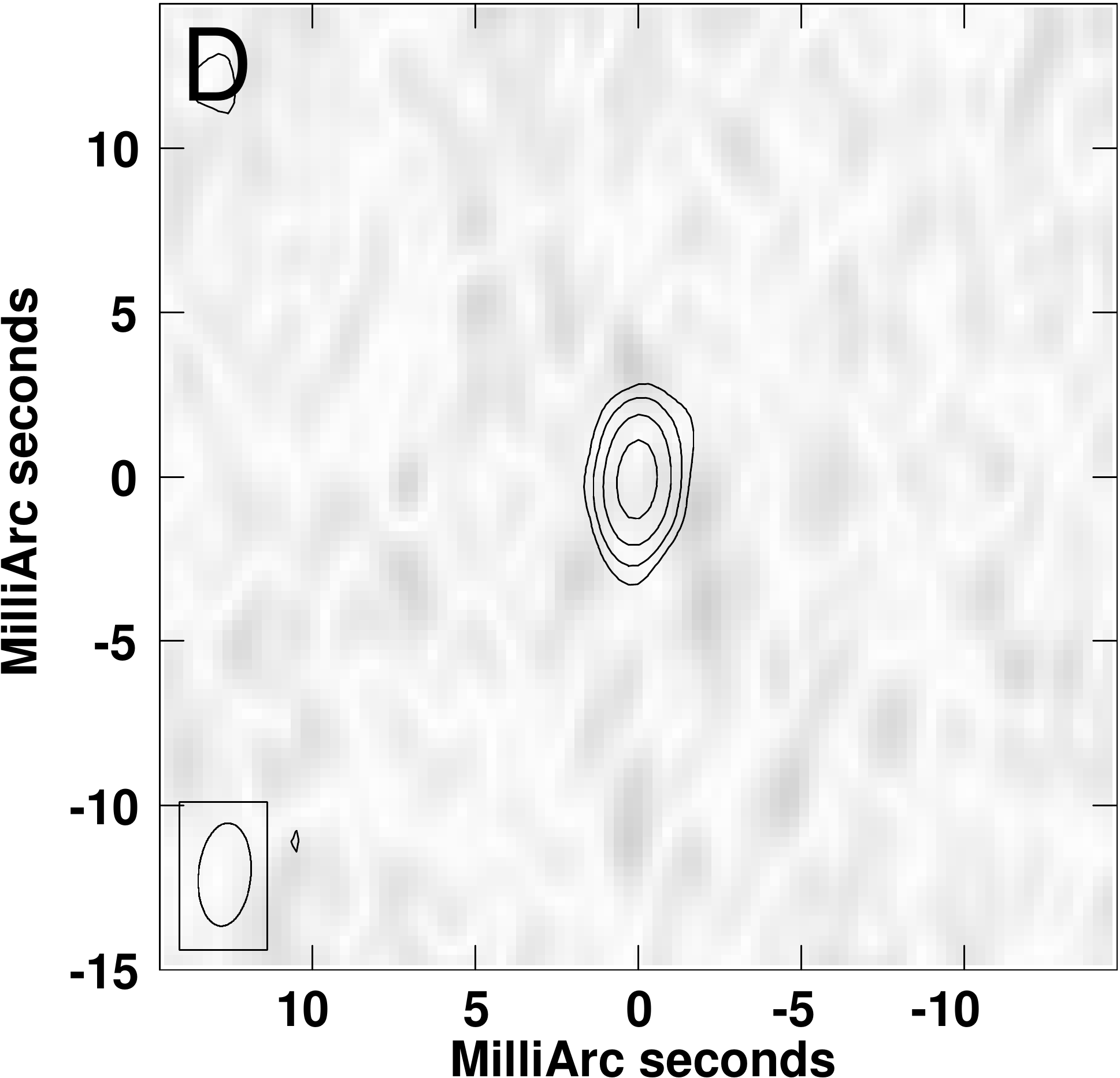}
    \includegraphics[scale=0.21]{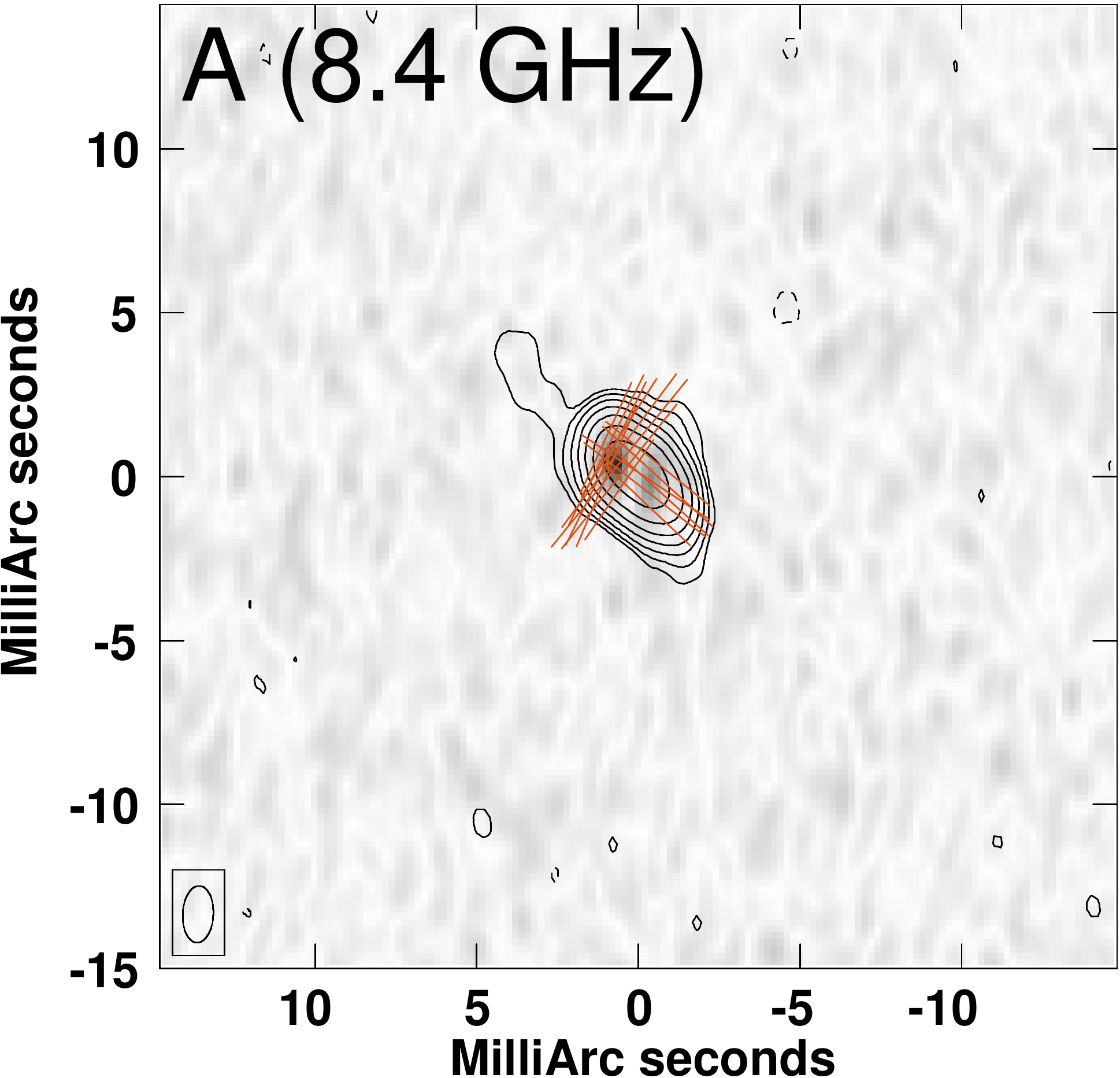}
    \includegraphics[scale=0.21]{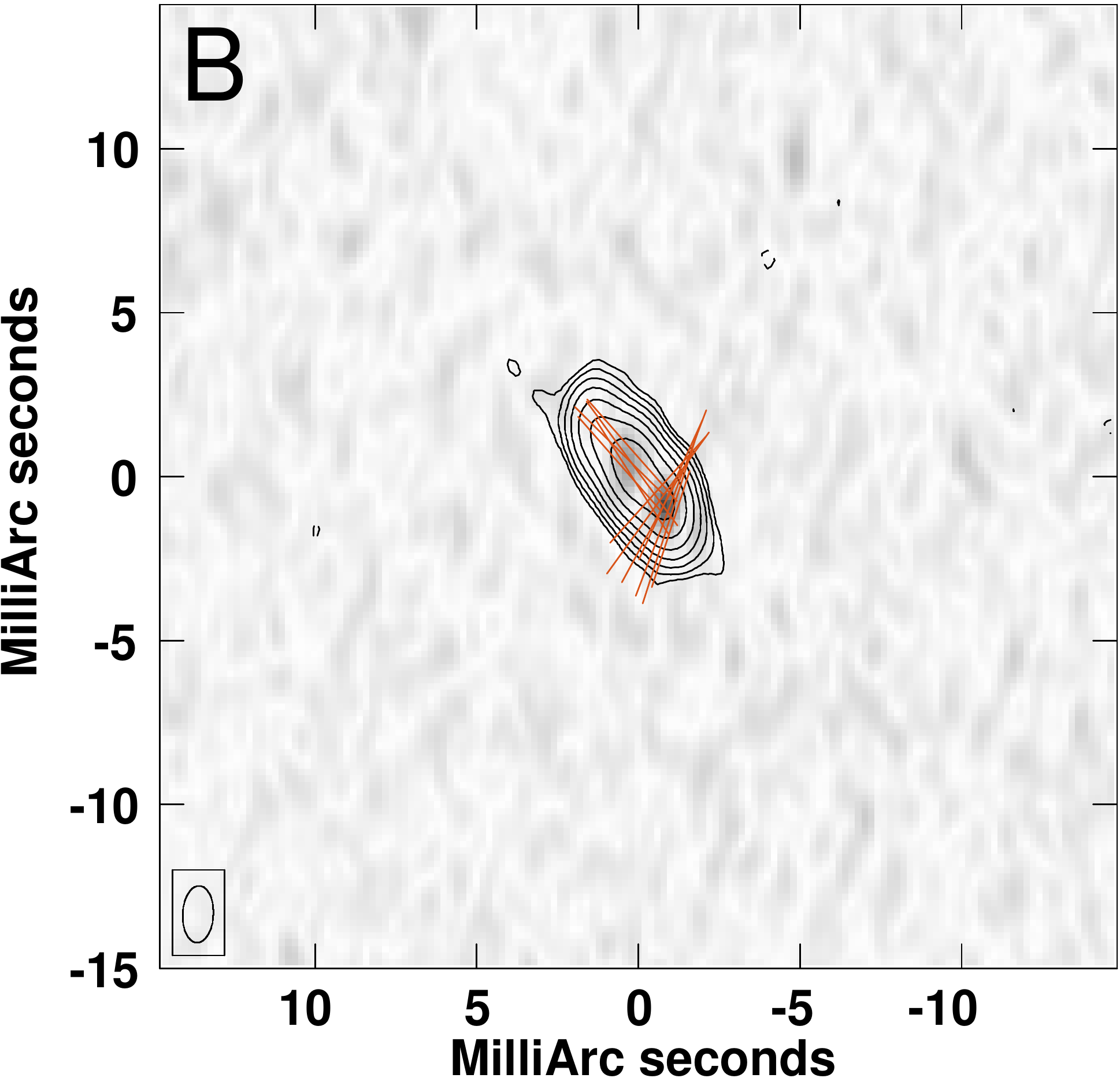}
    \includegraphics[scale=0.21]{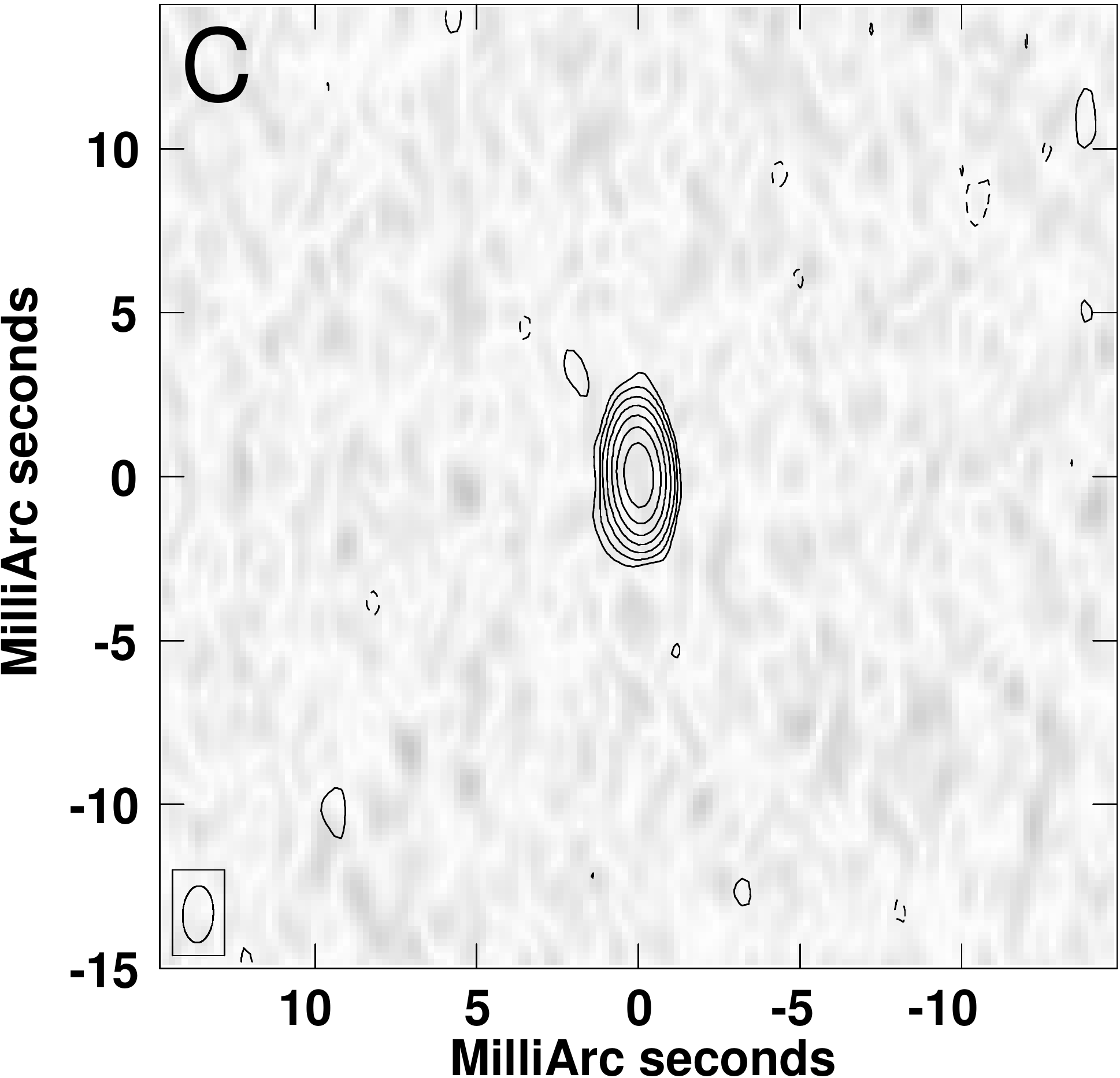}
    \includegraphics[scale=0.21]{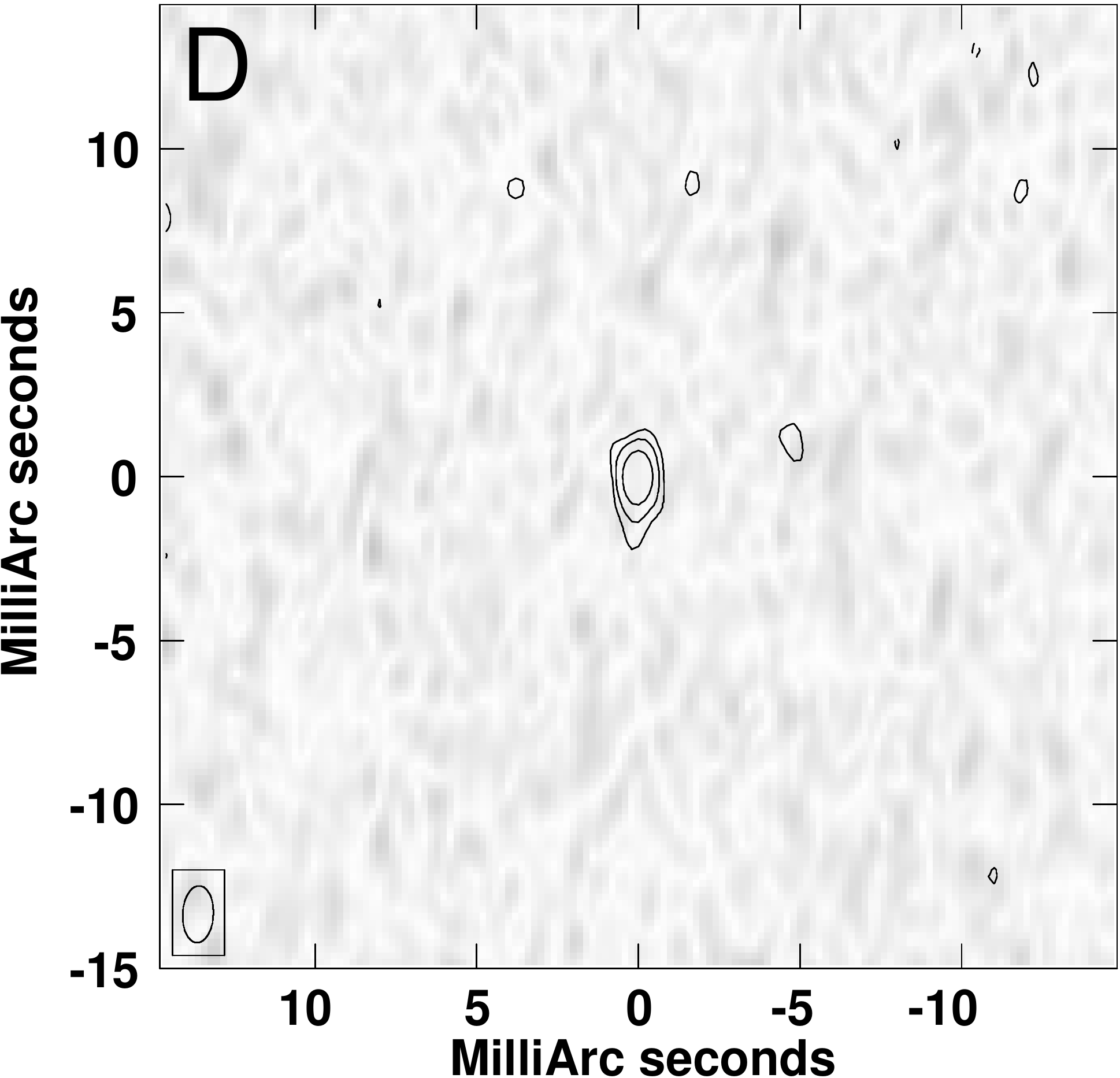}
    \caption{VLBA images of all four images of B1422+231 at 5 (top) and 8.4~GHz (bottom). Contours and grey scales represent the total intensity and linear polarization respectively. The restoring beam of each image is $3.1 \times 1.6$~mas$^2$ (5~GHz) and $1.6 \times 1.0$~mas$^2$ (8.4~GHz) and contours are plotted at $-1$, 1, 2, 4. 8, etc. multiples of three times the rms noise ($\sigma_5 = 150~\mu$Jy\,beam$^{-1}$ and $\sigma_{8.4} = 180~\mu$Jy\,beam$^{-1}$). All grey scales are plotted to a maximum of 1.5~mJy\,beam$^{-1}$. The angle of linear polarization is shown with sticks for pixels where the polarization is detected at a significance of at least 5~$\sigma$.}
    \label{fig:vlba}
  \end{center}
\end{figure*}

\begin{figure}
  \begin{center}
    \includegraphics[width=0.9\linewidth]{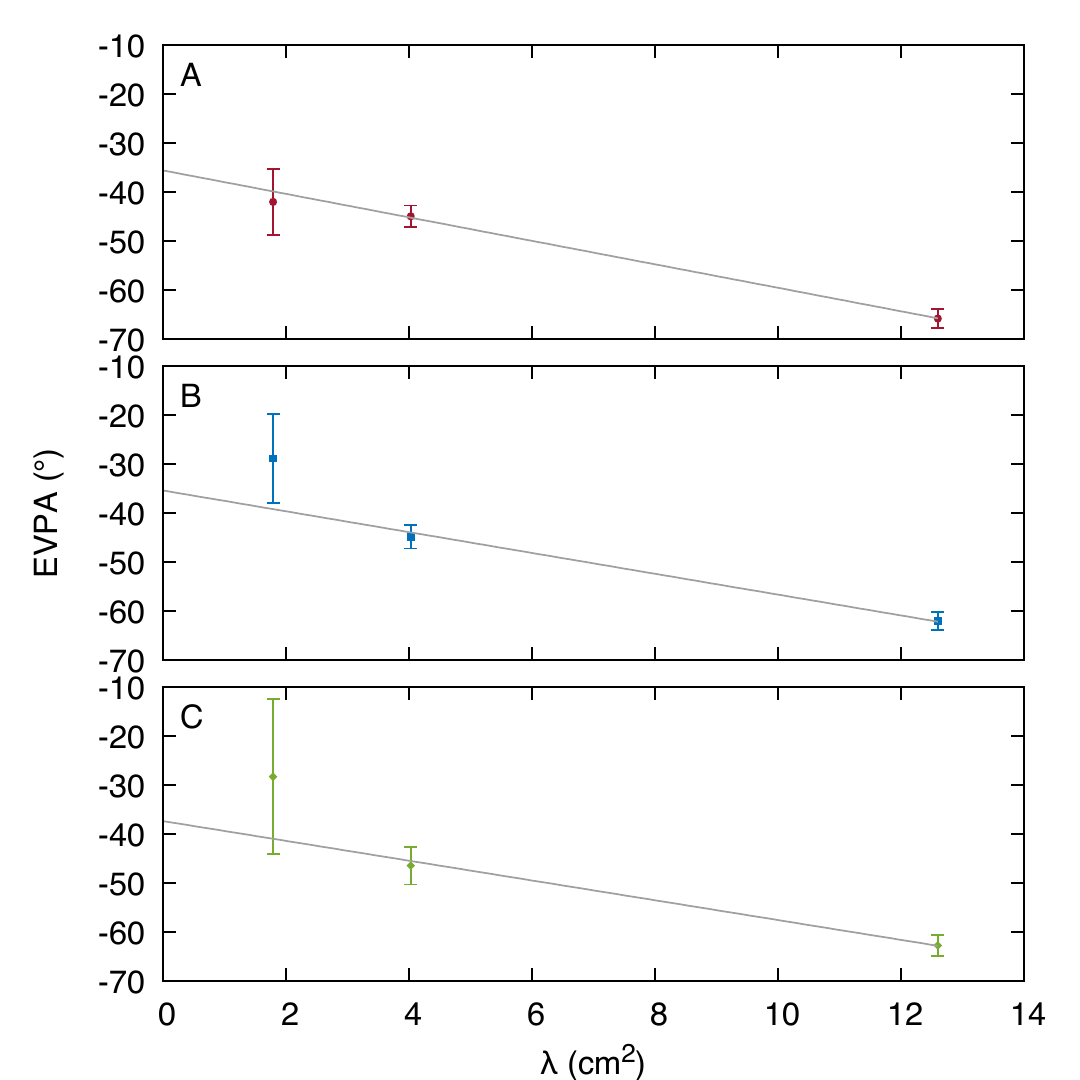}
    \caption{EVPA of images A, B and C as a function of $\lambda^2$ for data taken on 1994 February 22 at 8.4, 15 and 22~GHz. The slope of the fit gives the rotation measure and this and the measured intercept (intrinsic source EVPA) are the same for each image within the errors.}
    \label{fig:rm}
  \end{center}
\end{figure}

As demonstrated by \citet{biggs21}, a natural by-product of radio monitoring is an extremely high sensitivity map created by subtracting the fitted flux densities from the individual epochs and then combining the \textit{u,v} data from each. We remove the lensed emission using \textsc{uvmod} which allows separate subtraction of Stokes $I$, $Q$ and $U$ flux densities and then combine the epochs using the \textsc{stuffr} procedure. Removing the lensed images is important as combining data containing time-variable emission will lead to mapping artefacts.

A high-sensitivity map has been made by subtracting all four lensed images from the data and then combining the 97 8.4-GHz epochs from both seasons into a single dataset. This map, shown in Fig.~\ref{fig:stuffr}, has a sensitivity of 9.4~$\mu$Jy\,beam$^{-1}$ and allows the detection of a faint arc lying between B and C. The rms noise in a combined 15-GHz map is much higher, 50~$\mu$Jy\,beam$^{-1}$, and thus the arc is not detected. That this system would likely contain arcs was predicted by \citet{hogg94} and will likely originate from extended jet emission associated with the core of the lensed radio source.

As already mentioned, B1422+231 is a cusp lens i.e.\ the lensed source lies within the central tangential caustic, close to the point where two fold caustics meet \citep[see e.g.][]{narayan96}. We can draw some general conclusions about the source geometry by assuming a simple elliptically symmetric mass distribution in the lens galaxy. For example, as image~B lies closer to A than C, this indicates that the radio core lies closer to one of the folds. As the radio arc is instead dominated by emission located between images B and C, it therefore appears that the extended jet emission lies closer to and partially covers the \textit{other} fold caustic.

We have searched for the jet emission responsible for the arc using Very Large Baseline Array (VLBA) data taken in July 1995 as part of project BH010 (PI: J.~Hewitt). B1422+231 was observed at 5 and 8.4~GHz with eight six-minute scans at each frequency spread over a period of nine hours and resulting in relatively good \textit{u,v} coverage. The data were calibrated following standard procedures in \textsc{aips} with B1422+231 originally fringe-fitted with a simple model based on the VLA positions. The resultant maps (using a separate facet for each of the four images in \textsc{imagr}) were then used as an updated model for the fringe-fitting and this process repeated until the solutions converged. Multiple scans of 3C~84 were available for polarization D-term calibration and this was performed using \textsc{lpcal} after mapping and self-calibrating this source's data. Absolute EVPA calibration was done using a single scan of J1310+3220 with the EVPA of this source itself measured from VLA archival data taken within a few weeks of the VLBA observations.

The VLBA maps are shown in Fig.~\ref{fig:vlba}. The bright images are dominated by the moderately extended structures detected in previous VLBI observations \citep[e.g.][]{patnaik98,patnaik99}, but in images A and B there is a clear detection of an additional jet feature lying along the arc formed by A, B and C. This is particularly obvious at 5~GHz, whilst at 8.4~GHz the jet is only clearly detected in image~A. There are signs of the jet in image~C at 5~GHz, although this is more difficult to see due to its lying along the major axis of the restoring beam.

Linear polarization is also detected in all three bright images, but only at 5~GHz in image~C. A single polarized component is detected at the lower frequency, but at 8.4~GHz an additional component is visible. The EVPAs of the two components differ by approximately 90\degr, emphasizing how polarization variations at the angular resolution of the VLA could occur if the relative contribution of each were to change with time (Section~\ref{sec:extrinsic}). As expected, D appears unresolved and no polarization is detected due to a lack of sensitivity.

The jet emission revealed by our new VLA and VLBA maps could provide additional lens-model constraints, particularly if they could be mapped with better sensitivity. The rms noise of the VLBA data is relatively high, 150--180~$\mu$Jy\,beam$^{-1}$, an order of magnitude higher than what is possible with modern broad-band VLBI arrays. More modest improvements would be possible with the arcsec-scale mapping as the map presented here is already quite sensitive, but a higher-dynamic-range map of the arc along its whole length could be used to search for substructures such as those suggested as being responsible for the anomalous flux ratios in this system.

\subsection{The polarization properties of the lensed images}

\begin{table}
  \centering
  \caption{Rotation measures of the three bright images corresponding to the data and fits shown in Fig.~\ref{fig:rm}. The RMs are similar and consistent at a confidence level of 1~$\sigma$. The intercept of each fit, EVPA$_0$, measures the intrinsic EVPA in the absence of Faraday rotation and should be the same for each image, as is observed.}
  \label{tab:rm}
  \begin{tabular}{ccc} \\ \hline
    Image & RM (rad\,m$^{-2}$) & EVPA$_0$ ($^{\circ}$) \\ \hline
    A & $-418\pm20$ & $-35.6 \pm 1.1$ \\
    B & $-371\pm70$ & $-35.4 \pm 4.1$ \\
    C & $-352\pm72$ & $-37.3 \pm 4.6$ \\ \hline
  \end{tabular}
\end{table}

In a variation of the $u,v$ combination method, we have combined epochs solely from Season~1 i.e.\ when the polarization of the lensed source was relatively constant. The resulting 8.4-GHz map (with A, B and C subtracted) is also shown in Fig.~\ref{fig:stuffr} and detects image~D with a peak polarized flux density of 68~$\mu$Jy\,beam$^{-1}$, $1.7\pm0.1$~per~cent of the total intensity. This is similar to the magnitude of polarization in the other images, which have an average value of 1.5~per~cent during this season. The EVPA of image~D is $-50\pm2$\degr, significantly different to the other images which have average values between $-65$ and $-60$\degr. No detection of polarization was made at 15~GHz due to the higher noise at this frequency.

We have also used the single epoch of AP282 which included a scan at 22~GHz to measure the rotation measures (RM) of the three bright images. The 22-GHz data were observed in a very similar way to the other frequencies and 3C~286 was again used to calibrate the absolute angle of polarization. In doing this we used a recent version of \textsc{rldif} which takes into account the change in 3C~286's EVPA at high frequencies \citep{perley13b}.

The EVPA measurements of images A, B and C are shown in Fig.~\ref{fig:rm} together with linear fits to the data as a function of $\lambda^2$. A $\lambda^2$-variation of EVPA is a signature of Faraday rotation caused by passage of the radio waves through a magnetoionic plasma and is observed in many radio lenses. The slope of the fits gives the rotation measure (RM) and although this can be different for each image, the EVPA at zero wavelength (the value in the absence of Faraday rotation, EVPA$_0$) should be the same, which they they are. The fitted values of RM and EVPA$_0$ are given in Table~\ref{tab:rm}.

The RMs of the three bright images are consistent at a confidence level of 1~$\sigma$ which suggests that there is no differential RM caused by passage through either the lensing galaxy and/or the interstellar medium of the Milky Way. This is consistent with the lensing galaxy being an elliptical and the bulk of the RM is probably intrinsic to the lensed source, with a weighted average of $-410 \pm 19$~rad\,m$^{-2}$. However, the $\Delta$EVPA of up to 15\degr\ between D and the other images at 8.4~GHz suggests that this image \textit{is} subject to differential Faraday rotation, with a minimum differential rotation measure (assuming no $n\pi$ ambiguities) of $\Delta\mathrm{RM} \sim +200$~rad\,m$^{-2}$. However, this should be properly determined with high-sensitivity, multi-frequency observations and the discrepancy might be due to variability, either intrinsic or extrinsic.

Finally, our RM measurements are incompatible with the extremely high values ($-4230 < \mathrm{RM} < -3340$~rad\,m$^{-2}$) reported for the bright images by \citet{patnaik01} and we note that these authors did not find fits with a consistent EVPA$_0$ for each image.

\section{Conclusions}
\label{sec:conclusions}

We have presented a reanalysis of VLA monitoring data of the lens system JVAS~B1422+231 that were originally published by \citetalias{patnaik01b}, as well as unpublished data from another VLA monitoring campaign conducted a few years later. Using a \textit{u,v}-based modelfitting approach we have produced radio light curves of total flux density and polarization at 8.4 and 15~GHz.

We find significant intrinsic variability of the source, particularly in polarization during the second season, but there are few features in the variability curves that can be used for measuring a time delay due to the relatively long time-scales of the variability. Measuring the very short predicted time delays between the three bright images will always be difficult, but the nature of the intrinsic source variability makes this very difficult indeed. Ultimately, we do not believe that any of the existing VLA monitoring data are able to provide any strong constraints on the time delays in B1422+231.

The only realistic prospect for measuring a time delay in this system will be through future monitoring with modern high-sensitivity arrays such that the significantly fainter image~D is robustly detected at each epoch, in both total flux density and polarization. However, time-delay determination will be made more difficult due to the presence of extrinsic variability that causes each image to vary independently. As well as occurring on time-scales down to a few hours, these seem to particularly affect image~D and we suggest that this is due to microlensing of superluminal jet components by compact objects in the lensing galaxy, similar to what has been postulated for another lens system, CLASS~B1600+431. Future multi-frequency monitoring would allow a better characterisation of the extrinsic variability in this system.

The existing monitoring data have also allowed us to investigate other aspects of this system. Combining all epochs has led to the discovery of a faint arc joining images B and C at 8.4~GHz, and the jet that is presumably responsible for this has been detected using archival VLBA data. Higher sensitivity imaging of the arc could be used to search for dark-matter substructures in the lensing galaxy. Finally, we have measured reliable three-frequency rotation measures for the three bright images, the consistency of which argue for a RM that originates entirely in the lensed source. We have also detected the polarization of image~D for the first time and find its magnitude and EVPA to be similar to the other three.

\section*{Acknowledgements}

The author thanks Ian Browne for his continuing contributions to this series of papers and Tom Muxlow for his thoughts on the vagaries of VLBI data reduction. Thanks also goes to the anonymous referee whose report led to a number of improvements in the manuscript. The National Radio Astronomy Observatory is a facility of the National Science Foundation operated under cooperative agreement by Associated Universities, Inc.

\section*{Data Availability}

The data underlying this article will be shared on reasonable request to the corresponding author.



\bibliographystyle{mnras}
\bibliography{lensing}


\bsp	
\label{lastpage}
\end{document}